\documentstyle[12pt,bezier,axodraw]{article}
\newcommand{\bb}{\begin{equation}}
\newcommand{\ee}{\end{equation}}
\newcommand{\ba}{\begin{array}}
\newcommand{\ea}{\end{array}}
\newcommand{\beqa}{\begin{eqnarray}}
\newcommand{\eeqa}{\end{eqnarray}}

\newcommand{\as}{\alpha_{s}}

\newcommand{\ew}{electroweak }

\newcommand{\ms}{\overline{MS}}

\newcommand{\hff}{H\,\rightarrow\,f \bar{f}}
\newcommand{\cc}{coupling constant }
\newcommand{\ccs}{coupling constants }
\newcommand{\GB}{Goldstone boson }
\newcommand{\GBs}{Goldstone bosons }
\newcommand{\ra}{\rightarrow}
\newcommand{\rc}{renormalization constant }
\newcommand{\rcs}{renormalization constants }
\newcommand{\RC}{renormalization condition }
\newcommand{\RCs}{renormalization conditions }
\newcommand{\RSs}{renormalization schemes }
\newcommand{\RS}{renormalization scheme }
\newcommand{\ct}{counter term }
\newcommand{\cts}{counter terms }
\newcommand{\tp}{tadpole }
\newcommand{\tps}{tadpoles }
\newcommand{\pth}{perturbation theory }
\newcommand{\dZ}{\delta Z}
\newcommand{\dz}{\delta \zeta_v}
\newcommand{\BS}{broken symmetry }
\newcommand{\MS}{$\bar{MS}$ }
\newcommand{\DR}{dimensional regularization }
\newcommand{\lnM}{\ln(\frac{M^2}{\mu^2})}
\newcommand{\lnm}{\ln(\frac{m^2}{\mu^2})}
\newcommand{\yy}{{\cal Y}^2}
\title{\begin{flushright}
	{\normalsize PITT-TH-96-11 \\
 June 1996 \\}
	\end{flushright}
\vspace{0.2in}
{\bf Electroweak Parameters\\
 in the $\ms$ -Scheme \\}}
\author{{\bf A. I. Bochkarev}$^{a,\dagger}$ $\;$ and $\;$
{\bf R. S. Willey}$^{b}$ \\
{\it a TPI, U. of Minnesota, Minneapolis, MN 55455, USA}\\
{\it E-mail: bochkare@msi.umn.edu}\\
{\it b Physics Dept., U. of Pittsburgh, Pittsburgh PA 15260, USA}\\
{\it E-mail: willey@vms.cis.pitt.edu}}

\date{}

\begin{document}

\maketitle
%\vspace{0.3in}
\begin{center}
Abstract\\
\end{center}

We study electroweak parameters sensitive to the radiative
corrections, such as $\rho-ratio$ and $\hff$ decay 
rates in the $\ms$-scheme in the heavy $t$-quark mass 
limit $m_{t} \gg m_{w}$. In $\ms$-scheme the two-loop 
electroweak corrections $\,\sim \,m_{t}^{4}$
dominate over the QCD corrections $\,\sim\,\as m_{t}^{2}$
for $\rho$ . The relation
between the {\em on-shell} coupling constants and $\ms$-parameters is found 
to be rather sensitive to the Higgs boson mass. 
\vspace{0.2in}

\noindent $^{\dagger}$ On leave from: {\it Institute for Nuclear 
Research, Russian Academy of Sciences, Moscow 117312, Russia}

\vfill \eject

\section*{Introduction}
The standard model of \ew\ interactions is a renormalizable quantum field theory
. Thus one has the possibility of ``precision tests'' of the theory,and
sensitivity to ``new physics''.Two important issues are the scheme dependence
of the necessarily truncated perturbation series, and the decoupling or
nondecoupling of heavy masses from low energy processes --- in particular,
the dependence on the top and Higgs masses($m_t,m_H$).

The renormalization of the complete standard \ew\ theory is quite complicated
\cite{BSH}. There are all the complications of quantizing a nonabelian
gauge theory, as well as mass generation by spontaneous symmetry breaking,
$\gamma-Z^0$ mixing,etc. It is useful to find a substantially simpler
framework in which the issues raised above can be studied.

The top and Higgs masses enter the complete standard \ew\ theory through the
Yukawa coupling of the top and the Higgs and the quartic scalar self-coupling
and the vev of the (unshifted) higgs field.
\[
{\cal Y}_t
 \sim \frac{m_t}{v}, \hspace{1.in}  \lambda \sim \frac{m_H^2}{v^2}
\]
The simplified framework we will study is the ``gaugeless limit'' in which
the \ew\ gauge \ccs\ are set to zero
\[
g_2,g_1 \rightarrow 0
\]
and
\[
M_W,M_Z \sim gv \rightarrow 0
\]
Also, all Yukawa \ccs\  except $ {\cal Y}_t $  are set to $0$
\[
{\cal Y}_f \rightarrow 0 \hspace{.5in} (f \neq t)
\]
Keeping only the heavy quark generation,we study the scalar - (heavy) quark
sector, consisting of a physical Higgs, three unphysical \GBs\, and the t and b
quarks.This limit has no gauge degrees of freedom, gauge fixing, or Fadeev-
Popov factor. It has a global $SU(2)_L$ symmetry and spontaneous symmetry
breaking with three massless \GBs. (This treatment has substantial overlap with,
 but is not identical to, considerations based on the ``Equivalence Theorem''
\cite{Ri}.)

In this much simpler model, we will study in some detail the issues raised
in the first paragraph. In the end, to make contact with observed low energy
processes, we will find that we can not completely avoid the complications
of the renormalization of the full theory.

\section{Renormalization}

The fields of the reduced theory occur as singlets and doublets under the 
global $ SU(2)_L $.
\bb
\Psi_l = \left( \begin{array}{l}
\cal T \\
\cal B
\end{array} \right)_L, \; {\cal T}_R, \; \;{\cal B}_R, \; \left( \begin{array}{l}
\Phi_+ \\
\Phi_0 
\end{array} \right)
 \label{fields}
\ee
\medskip
\[
\Phi_+ = \varphi_+ \equiv \varphi = \frac{\varphi_1 - i \varphi_2}{\sqrt{2}},
\hspace*{.5in} \Phi_0 = \frac{{\cal H} - i \varphi_0}{\sqrt{2}} 
\] 
\medskip
\[
\tilde{\Phi} = i{\tau}_2 {\Phi}^{\dagger} = \left( \begin{array}{l}
\bar{\Phi_0} \\
- \Phi_-
\end{array} \right)
\]  

The Lagrangian is 
\beqa
{\cal L} & = & \bar{\Psi_L} i \gamma \cdot \partial \Psi_L + \bar{{\cal T}_R} 
i \gamma \cdot \partial {\cal T}_R + \bar{{\cal B}_R} i \gamma \cdot \partial 
{\cal B}_R  \nonumber \\
 & & + \partial \Phi^{\dagger} \partial \Phi - \mu^2_0 \Phi^{\dagger} \Phi   
     - \lambda_0 (\Phi^{\dagger}\Phi)^2 \nonumber\\ 
& & - {\cal Y}_{0_t}(\bar{\Psi_L} \tilde{\Phi} {\cal T}_R + h.c.) \label{Lag}   
\eeqa
The fields are unrenormalized canonical fields, and  $\mu^2_0$,$\lambda_0$,
and ${\cal Y}_{0_t}$ are bare mass and \ccs. 

To implement \pth in the \BS phase $(\mu^2_0< 0,\langle {\cal H} \rangle \neq 0 )$,
 the Higgs field is shifted
\bb
\langle {\cal H} \rangle = {\cal V}, \hspace{.5in} {\cal H}={\cal V}+ \hat{
\cal H}   \label{shift h} 
\ee
The condition that one is perturbing about the correct vacuum is
\bb
\langle \hat{\cal H} \rangle = 0   \label{vev=0}
\ee
This condition determines ${\cal V}$ as a function of the parameters of the 
theory. 

The bare fields and parameters of (\ref{fields}),(\ref{Lag}) are reexpressed 
as renormalized fields and parameters, multiplied by appropriate $Z$-factors. 
\beqa
\mu^2_0 = Z_{\mu} \mu^2, & \lambda_0 = Z_{\lambda} \lambda, & {{\cal Y}_0}_t = 
Z_y {\cal Y}_t  \nonumber \\
\Psi_L = \sqrt{Z_L} \psi_L, & {\cal T_R} = \sqrt{Z_t} t_R & {\cal B_R} = 
\sqrt{Z_b} b_R  \nonumber  \\
\{ \hat{\cal H},\varphi_0,\varphi,\varphi^{\dagger} \} = & \sqrt{Z_{\phi}} \{ 
h,\phi_0,\phi,\phi^{\dagger}\}, & {\cal V} = \sqrt{Z_{\phi}}V \label{renorm}    
\eeqa
In terms of these \rc, the Yukawa and quartic scalar vertex renormalizations 
are
\bb
\sqrt{Z_L}\sqrt{Z_t}\sqrt{Z_{\phi}}Z_y = Z_3, \hspace{.5in} Z^2_{\phi} 
Z_{\lambda}=Z_4  \label{vertrenorm} 
\ee  
Note that we have introduced a common field strength \rc for all four of the 
scalar fields. This is in accord with the dictat to introduce only \cts which
respect the symmetries of the original ``bare'' Lagrangian, in this case, the
$O(4)$ symmetry of the purely bosonic part of the Lagrangian (\ref{Lag}).  
Then after
 spontaneous symmetry breaking, the currents associated with the $O(4)$ 
generators will still be conserved. The fact that there is no longer a 
common mass shell will have to be taken into account in the LSZ reduction 
formulas relating Green functions to S-Matrix elements (see section 3).

When (\ref{shift h}) and (\ref{renorm}) are substituted into (\ref{Lag}), 
and all Z's are rewritten as $Z = 1 + \delta Z$,
the Lagrangian (\ref{Lag}) is rewritten as a lengthy sum of terms, starting 
with a sum of terms of the same form as (\ref{Lag}) but with all bare fields 
and parameters replaced by renormalized ones, plus terms proportional to $V$ 
(and containing no $\delta Z$), plus \cts proportional to one or more $\delta 
Z's$. The procedure is quite standard, and we will only write out individual 
pieces as we need to discuss them. 

The first order of business is to implement the stability condition (\ref{vev=0})
 To do this, compute all \tp graphs, including \cts, to a given order 
(number of loops) in \pth, and expand the renormalized $V$
\beqa
V \;\;&\equiv&\;\;{\zeta}_v \;v  \;,\nonumber \\
{\zeta}_v\;\;&=&\;\;1\;+\;{\zeta}_v^{(1)}\;+\;{\zeta}_v^{(2)}\;+\;\ldots 
 \label{zeta}
\eeqa
 and adjust ${\zeta}_v$ (equivalently $V$) to satisfy (\ref{vev=0})  order by 
order. The terms in $\cal L $, linear in the shifted Higgs field, are
\beqa
{\cal L}_1\;\,&=&\;\,-\,{\cal V}\,\hat{\cal H}\,(\mu_0^2 + \lambda_0 {\cal V}^2)\;\,=\;\, 
 -\,V\,h\,(Z_{\phi} Z_{\mu} \mu^2 + Z_{\phi}^2 Z_{\lambda} \lambda V^2) \nonumber \\
   &=&\;\;-\,v\,h\,\{(\mu^2 + \lambda v^2) + (\zeta_v-1)(\mu^2 + \lambda v^2) + 
  [(Z_{\phi} Z_{\mu}-1)\mu^2 + (Z_4\zeta_v^2 -1)\lambda v^2]\} \nonumber  \\ 
  &+&\; \mbox{two-loop counterterms}  \label{L1} 
\eeqa 
Zeroth order fixes
\bb
 v^2 = \frac{- \mu^2}{\lambda}, \hspace{.5in}  (\mu^2 = - \frac{M^2}{2}<0) 
\label{V0}  
\ee
and the one-loop \ct may be written as 
\beqa
  {\cal L}_1 \;\;&=&\;\; v\, \delta \mu^2\, h \nonumber  \\
  \delta \mu^2\;\;&=&\;\; -\,(Z_{\phi} Z_{\mu} - Z_4 \zeta_v^2)\,\mu^2 \;\;=\;\;
 \left(\,Z_{\phi} Z_{\mu} \;-\; Z_4 \zeta_v^2\,\right) \,\lambda\, v^2.
 \label{delmu} 
\eeqa
Computation of $\langle h \rangle $ in one-loop order (fig. 1) gives

\begin{center}\begin{picture}(440,160)(0,0)

\SetScale{.8}
\DashCArc(50,110)(40,0,360){3}
\DashLine(50,10)(50,70){6}
\Text(40,135)[b]{$\large \phi_{\stackrel{+}{-}},\phi_0$}

\DashCArc(170,110)(40,0,360){6}
\DashLine(170,10)(170,70){6}
\Text(135,135)[b]{\Large h}

\CArc(290,110)(40,0,360)
\DashLine(290,10)(290,70){6}
\Text(240,135)[b]{\Large t}

\CArc(410,110)(5,0,360)
\DashLine(410,10)(410,105){6}
\Text(330,90)[]{\Large x}

\end{picture}\\
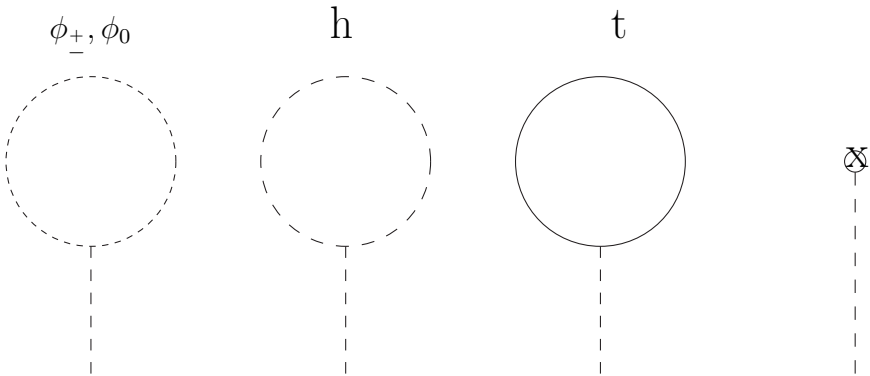
Fig.1.Tad-pole diagrams.Long dashes - Higgs, short dashes - Goldstone,
    solid - top. Big dot with x is counter term linear in Higgs field.
\end{center}

\bb
 \langle h \rangle \;\;=\;\; \frac{-1}{M^2}\;\left[\;2 \lambda v A_0\;+\;
\lambda v A_0 \;+\;3\lambda v A_M \;-\; 2 \sqrt{2} N_c {\cal Y} m A_m\; -\;
 v \delta \mu^2 \;\right]  \label{V1} 
\ee
where the regularized Feynman integral is
\bb
 A_M \;\;=\;\; i \int_{reg}(dp)\;\frac{1}{p^2-M^2+i\epsilon}   \label{AM}
\ee 
(In the reduced theory there are only two masses, $m_t = m$ and $m_H = M$. The 
\GBs remain massless). 
Expanding (\ref{delmu}) to one-loop order, requiring (\ref{V1}) to be zero,
fixes ($\sqrt{2} m = {\cal Y} \; v $)
\bb
 \delta \mu^2\;\; = \;\;\frac{M^2}{2}\;\left(\,\dZ_{\phi}+ \dZ_{\mu} - \dZ_4 -
 2 \dz\,\right)\;\;=\;\;3 \lambda (A_0 + A_M) \;-\; 2 N_c y^2 A_m \label{delmu=} 
\ee
We will subsequently fix $\dZ_{\phi}$ as the \GB field \rc and $\dZ_4$ as the 
Higgs boson mass \rc ($M^2 = 2 \lambda v^2$). This leaves $\dZ_{\mu}$ and 
$\dz$ to be adjusted to satisfy (\ref{delmu=}). At this point we find a  
significant difference between momentum dependent (``$MOM$'') 
 renormalization schemes, and momentum independent, in particular 
\MS, renormalization schemes. In an $MOM$ scheme, $ \dZ_{\mu}$ is unrestricted 
so we can choose $\dz$ to be zero and adjust $\dZ_{\mu}$ to satisfy 
(\ref{delmu=}). In \MS, $\dZ_{\mu}$ is restricted to a divergent part;  no 
adjustable finite part is admitted. Thus when we compute the integrals on 
the right hand side of (\ref{delmu=}) with \DR the $\frac{1}{\epsilon}$ terms 
are matched by $\frac{1}{\epsilon}$ terms from the $\delta Z's $ on the left 
hand side, but there is a finite part remaining which determines $\dz$. In 
\DR 
\beqa
A_M\;\;&=&\;\;\frac{M^2}{16{\pi}^2} \left[\,-\, \Delta_{\epsilon}\,+\,
\ln(\frac{M^2}{\mu^2})\,-\,1\,\right] \nonumber \\  
\Delta_{\epsilon}\;\;&=&\;\;\frac{2}{4-d} \,-\, \gamma_E + \ln(4 \pi)   \label{AMDR} 
\eeqa
$\mu^2$ is the arbitrary scale introduced in \DR, not to be confused with 
the arbitrary renormalized mass parameter in (\ref{renorm}), (\ref{V0}).     
 Then
\bb
\bar{\dz}\;\; =\;\; \frac{1}{16 \pi^2}\;\left[\,3\,\lambda\,\left(1 - \lnM \right)
\;-\;2 N_c\,{\cal Y}^2\,\frac{m^2}{M^2}\,\left(1 - \lnm \right) \,\right] 
\label{zetabar}  
\ee 
Note that the right hand side of (\ref{delmu=}) depends on the regularization 
of the divergent integrals, but makes no reference to the renormalization 
scheme. The combination of \rc plus the one-loop shift in the v.e.v. occuring
in $\delta \mu^2$ is scheme independent.  

The terms in the Lagrangian, quadratic in the scalar fields are 
\beqa 
{\cal L}_2 \;\;=\;\; \partial \Phi^{\dagger} \partial \Phi\;-\; \frac{1}{2}
(\mu_0^2 + \lambda_0 {\cal V}^2)(2 \varphi^{\dagger} 
\varphi + \varphi_0^2 ) \;-\; \frac{1}{2}(\mu_0^2 + 3 \lambda_0 {\cal V}^2) h^2 \nonumber \\   
\;\;= \;\;\partial \Phi^{\dagger} \partial \Phi\; +\; \frac{\delta \mu^2}{2}
(2 \phi^{\dagger} \phi + \phi_0^2) - 
\frac{1}{2} (2 \lambda v^2)h^2 + \frac{1}{2}(\delta \mu^2 - 
(Z_4 \zeta_v^2 -1)(2 \lambda v^2))h^2     \label{L2s} 
\eeqa
 \bb 
  M_{\phi} = 0,  \hspace{.5in}  M^2 = 2 \lambda v^2  \label{bosmass}  
\ee 
The one-loop \cts (for $-i \Sigma$) generated by (\ref{L2s}) are 
\beqa
&\phi& : \;\;\; i(Z_{\phi} -1)q^2 + i \delta \mu^2  \\
&h& : \;\;\; i(Z_{\phi} -1)q^2 + i \delta \mu^2 -i (Z_4 \zeta_v^2 -1)M^2 \label{ct2s} 
\eeqa 
The one-loop \GB self-energy diagrams, including \tps and \cts are shown in 
fig. 2.

\begin{center} \begin{picture}(440,230)(0,0)
\SetScale{.8}

\DashCArc(40,200)(20,0,360){3}
\DashLine(10,180)(40,180){3}
\DashLine(40,180)(70,180){3}

\DashCArc(115,200)(20,0,360){6}
\DashLine(85,180)(115,180){3}
\DashLine(115,180)(145,180){3}

\DashCArc(200,180)(20,0,180){3}
\DashCArc(200,180)(20,180,360){6}
\DashLine(160,180)(180,180){3}
\DashLine(220,180)(240,180){3}

\CArc(295,180)(20,0,360)
\DashLine(255,180)(275,180){3}
\DashLine(315,180)(335,180){3}

\CArc(380,180)(5,0,360)
\DashLine(350,180)(375,180){3}
\DashLine(385,180)(410,180){3}
\Text(306,142)[]{\Large x}

\DashCArc(80,70)(20,0,360){3}
\DashLine(80,10)(80,50){6}
\DashLine(50,10)(80,10){3}
\DashLine(80,10)(110,10){3}

\DashCArc(160,70)(20,0,360){6}
\DashLine(160,10)(160,50){6}
\DashLine(130,10)(160,10){3}
\DashLine(160,10)(190,10){3}

\CArc(240,70)(20,0,360)
\DashLine(240,10)(240,50){6}
\DashLine(210,10)(240,10){3}
\DashLine(240,10)(270,10){3}

\CArc(320,70)(5,0,360)
\DashLine(320,10)(320,65){6}
\DashLine(290,10)(320,10){3}
\DashLine(320,10)(350,10){3}
\Text(259,55)[]{\Large x}

\end{picture}\\ 
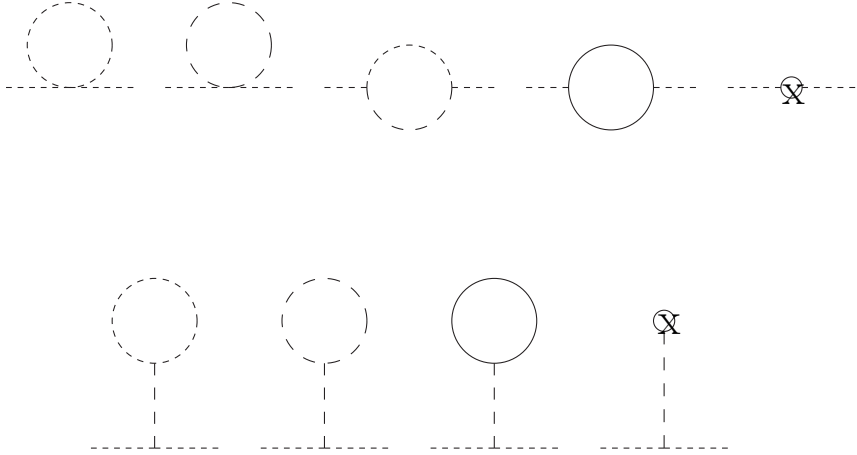
Fig.2. The one-loop self-energy diagrams of the Goldstone field.
\end{center}

 The \tp diagrams plus the \tp \ct in the second line of fig. 2 add to 
zero by our previous choice of $\delta \mu^2$ (\ref{delmu=}). The contribution 
of the Feynman diagrams and \ct of the first line of fig. 2 to the renormalized 
self-energy function is, for $\Sigma_+ $
\beqa
 - i\Sigma_{\phi_+} = -i 2 \lambda[2 A_0 + \frac{1}{2}A_0 +\frac{1}{2} A_M + 
 2 \lambda v^2 I_{M0}(q^2)] \nonumber  \\
 +i {\cal Y}^2 N_c [A_0(q^2) +A_m(0) -(q^2 -m^2)I_{m0}(q^2)] +i \delta \mu^2 + 
 i\dZ_{\phi} q^2  \label{sigma+} 
\eeqa 
and for $\Sigma_0$ 
 \beqa
   - i\Sigma_{\phi_0} = -i 2 \lambda[ A_0 + \frac{3}{2}A_0 +\frac{1}{2} A_M +
  2 \lambda v^2 I_{M0}(q^2)]  \nonumber \\
   +i \yy  N_c [A_m(q^2) +A_m(0) -q^2 I_{mm}(q^2)] +i \delta \mu^2 + 
  i\dZ_{\phi} q^2   \label{sigma0}
\eeqa
Here 
\bb  
 A_m(q^2) = i \int_{reg}(dl)\frac{1}{(l-q)^2 - m^2}, \hspace{.5in} 
I_{a b}(q^2)  = i \int_{reg}(dl)\frac{1}{(l^2 - a^2)((l-q)^2 - b^2)}  \label{AI1} 
\ee
and 
\bb
 A_m(0) = A_0(0) + m^2 I_{m 0}(0)   \label{AI2}  
\ee 
for any regularization. Using (\ref{AI2}) and (\ref{delmu=}) in (\ref{sigma+},
(\ref{sigma0}), we find
\bb
 \Sigma_{\phi_+}(0) = \Sigma_{\phi_0}(0) = 0    \label{mphi} 
\ee
i.e. the same $\delta \mu^2$ which enforces $\langle h \rangle = 0 $ also 
makes $M_{\phi}=0$ for both charged and neutral \GBs\.This is the reason that it is possible to choose $\dz$ equal to zero in an MOM scheme.  It is interesting to 
add the two \ct contributions from the first and second lines of fig. 2.
\[
 i \delta \mu^2 \;+\; (- 2 i \lambda v) \,\frac{i}{- M^2}\,i v\,\delta \mu^2 
\;\;=\;\; i \delta \mu^2 \,(1 \;-\; \frac{2 \lambda v^2}{M^2}) \;\;=\;\; 0
\] 
Thus, the sum of all the Feynman diagrams, including the \tp contributions, 
but no \cts\, also gives zero for $q^2=0$. This is a consistency check, that 
the Goldstone theorem is satisfied independently of renormalization scheme.

The $\dZ_{\phi}$ \ct is determined by the terms linear in $q^2$ in (\ref{sigma+}) , 
(\ref{sigma0})
\bb 
 {\Sigma}'_{\phi_+}(0) = 2 \lambda[M^2 I'_{M0}(0)] - \yy N_c [A'_0(0) -
 I_{m0}(0) + m^2 I'_{m0}(0)] - \dZ_{\phi}  \label{sig+} 
\ee 
\bb 
  {\Sigma}'_{\phi_0}(0) = 2 \lambda[M^2 I'_{M0}(0)] - \yy N_c [A'_m(0) -
 I_{mm}(0) ] - \dZ_{\phi}  \label{sig0}  
\ee 
In an on-shell $MOM$ renormalization scheme, the field strength \rc $\dZ_{\phi}$ 
is fixed  such that the residue of the pole of the renormalized $\phi$ propagator 
is unity. In the spontaneously broken symmetry phase, this would require 
separate $\dZ_{\phi_+}$ and $\dZ_{\phi_0}$. But then the renormalization 
reparametrization $(\ref{renorm})$ of the Lagrangian would introduce explicitly 
$SU(2)_L$ breaking terms into the Lagrangian $(\ref{Lag})$. So we stick with a 
single $\dZ_{\phi}$ chosen ,for the moment arbitrarily, to be $\dZ_{\phi_+}$. 
However, $\dZ_{\phi_0}$ will also be required for later use in the LSZ 
reduction formulas. If the integrals in (\ref{sig+}),(\ref{sig0}) are computed 
with \DR, the results are 
\bb
 \dZ_{\phi} \equiv \dZ_{\phi_+} = \frac{\lambda}{16 \pi^2}[-1] + \frac{\yy}
{16 \pi^2} N_c [- \Delta_{\epsilon} + \lnm - \frac{1}{2}]  \label{dz+} 
\ee 
\bb
 \dZ_{\phi_0} = \frac{\lambda}{16 \pi^2}[-1] + \frac{\yy}
{16 \pi^2} N_c [- \Delta_{\epsilon} + \lnm ]  \label{dz0} 
\ee 
We remark here that in a treatment of the entire \ew theory, $\delta Z_{\phi}$  may be fixed by a Renormalization condition in the gauge sector. We will return to this 
point when we discuss a Ward-Slavnov-Taylor identity of the theory. 
The $\ms$ field strength renormalization is the same for $\phi_+$ and $\phi_0$.
It is simply the common $\Delta_{\epsilon}$ contribution. 
\bb 
 \bar{\dZ_{\phi}}\;\;=\;\; N_c\;\frac{\yy}{16 \pi^2}\;
\left[\,- \Delta_{\epsilon} \,\right]     \label{dzbar} \ee 
 
For the physical Higgs propagator, there is a set of self-energy Feynman 
diagrams corresponding to those of fig. 2 for the \GBs. The resulting 
renormalized self-energy function is  
\beqa
-i\Sigma_h(q^2) \;\;=\;\; &-&\;2 i \lambda\;\left[\,\frac{3}{2}A_0\;+\;
\frac{3}{2}A_M\;+\;2 \lambda v^2\,\left(\,\frac{3}{2}I_{00}(q^2) \,+\,
\frac{9}{2}I_{MM}(q^2)\,\right)  \,\right] \nonumber  \\
&+&\;i\yy\; N_c\; \left[\,A_m(q^2)\;+\;A_m(0)\;-\;(q^2 - 4 m^2)I_{mm}(q^2)\,\right]
\nonumber  \\
&+&\;i\dZ_{\phi} q^2 \;+\;i\delta \mu^2\;-\;i(Z_4 \zeta_v^2 -1)\,M^2 \label{sigmah} 
\eeqa 
The on-shell mass \RC for an unstable particle is generally taken to be   
$Re \Sigma(M^2)=0$. 
Substitute $\delta \mu^2$ from (\ref{delmu=})(recall that it is scheme 
independent) into (\ref{sigmah}), and set $q^2=M^2$. (With \DR $A_m(q^2)=
A_m(0) \equiv A_m$).
\beqa
\Sigma_h(M^2)\;\;&=&\;\;3\lambda\,M^2\,\left[\,I_{00}(M^2)\,+\,3I_{MM}(M^2)\,
\right] \nonumber \\
&+&\; N_c \,\yy\, \left[\,(M^2 - 4 m^2)I_{mm}(M^2)\,\right]  
\;-\;\dZ_{\phi} M^2 \;+\; 2 \dz M^2 \;+\;\dZ_4 M^2   \label{sigM} 
\eeqa
Since $\dZ_{\phi}$ and $\dz$ are previously fixed, in either $MOM$ or \MS,
taking the real part of $I_{00}(M^2)$ (and $I_{mm}(M^2)$ if $M > 2 m$) and 
setting Re$\Sigma_h(M^2)$ to zero fixes $\dZ_4$ in $MOM$, while just matching 
$\Delta_{\epsilon}$ terms fixes $\dZ_4$ in \MS.

Taking the derivative of (\ref{sigmah}) gives
\beqa
 {\Sigma}'_h (M^2)\;\;&=&\;\; 3 \lambda M^2[I'_{00}(M^2)\,+\,3 I'_{MM}(M^2)] 
\nonumber \\
-N_c\,\yy\,\left[\,-I_{mm}(M^2)\,-\,(M^2 - 4 m^2)I'_{mm}(M^2)\,\right]\; -\; 
\dZ_{\phi}.  \label{sigh} 
\eeqa
The fact that this is not zero for $\dZ_{\phi}$ of (\ref{dz+}) has  
implications for calculation of processes in which a physical Higgs particle 
appears as an external line. When $\Sigma'_h(M^2)$ is not zero, virtual 
radiative corrections to external Higgs lines survive LSZ amputation. These 
contributions can be determined from the difference $\dZ_h - \dZ_{\phi}$. 
This is simple in the limit $M^2 \gg m^2$ ($\lambda \gg \yy$).
\bb
 \dZ_h = \frac{\lambda}{16 \pi^2}[3 + 9(1-\frac{2 \pi}{3 \sqrt{3}}) + {\cal O}
(\frac{m^2}{M^2})]   \label{dzh} 
\ee
Comparison with (\ref{dz+}) gives
\bb
 \dZ_h - \dZ_{\phi} = \frac{\lambda}{16 \pi^2}[13 - 2\sqrt{3} \pi + 
      {\cal O}(\frac{m^2}{M^2})]  \label{diffz} 
\ee

From consideration of the bosonic one- and two- point functions we have fixed 
the one-loop bosonic \rcs, $\dZ_{\phi},\dZ_{\mu},\dZ_4$, and the one-loop 
shift of the vev, $\dz$,in both MOM and \MS schemes. There remain bosonic 
three- and four-point functions with ultraviolet divergences, but a sequence 
of Ward identities (e.g. \cite{JW}) guarantee that these will be rendered 
finite by  the \cts generated 
in (\ref{Lag}) by (\ref{shift h}) and (\ref{renorm}), (\ref{vertrenorm}). 

We turn now to the relation between the bosonic $MOM$  mass and \cc  
$(M^*,\lambda^*)$ and the \MS mass and \cc ($\bar{M},\bar{\lambda}$).    
The OS mass is determined by the mass shell condition
\bb 
  0 =  \mbox{Re}D_h^{*^{-1}}(M^{*^2}) = (M^{*^2}) - (M^{*^2}) -\mbox{Re}
    \Sigma_h^* (M^{*^2})   \label{D-1} 
\ee
This fixes $\dZ_4^*$ in (\ref{sigM}). The bosonic $MOM$ parameters are $M^*$ 
and $\lambda^*$. $M^*$ is the physical Higgs mass (modulo the usual problems 
of unstable particles).  Since $\dZ_4^*$ is fixed by the mass condition, it is 
not available to define the \cc  as the value of a vertex function at 
some kinematic point. $\lambda^*$ is traded for $v^*$ which is to be 
determined in terms of the accurately known low energy \ew parameters $G_F,
\alpha,M_Z$. Thus 
\bb
 \lambda^* = \frac{M^{*^2}}{2 v^{*^2}}   \label{lamstar} 
\ee
In \MS Eqn. (\ref{D-1}) becomes   
\bb
  0 =  \mbox{Re}\bar{D}_h^{-1}(M^{*2}) = M^{*2} - \bar{M}^2 -\mbox{Re}
   \bar{ \Sigma}_h (M^{*^2})   \label{D-1MS}
\ee
Then
\beqa
\frac{M^{*2}}{\bar{M}^2} \;\;&=&\;\; 1 \;+\;\mbox{Re}\bar{\Sigma}_{h} (M^{*2})/\,\bar{M}^2 
\nonumber  \\ 
  &=& \;\;1 \;+\; 3 \lambda[\mbox{Re}I_{00}(M^2)+3 I_{MM}(M^2)] \nonumber \\
&+&\;\;N_c \yy[(1-4\frac{m^2}{M^2})\mbox{Re}I_{mm}(M^2)] 
\;-\;\bar{\dZ}_{\phi}\;+\;\bar{\dZ}_4\; +\; 2 \bar{\dz}   \label{MM}  
\eeqa
due to (\ref{sigM}). $\bar{\dZ_{\phi}}$ and $\bar{\dZ_4}$ just remove the 
$\Delta_{\epsilon}$ terms. Then 
\bb
  \frac{M^{*^2}}{\bar{M}^2} = 1 + \frac{\lambda}{16 \pi^2}[12 \lnM -24 + 
3\sqrt{3}\pi +{\cal O}(\frac{m^2}{M^2})] + 2 \bar{\dz} 
   \label{MMM}  
\ee
The \MS mass depends on the \MS one-loop shift of the v.e.v., $\bar{\dz}$, given  
in (\ref{zetabar}). Note that an alternative definition of the $\ms$ mass is       possible. If (\ref{MMM}) is multiplied through by $\bar{M}^2$, the collected    terms $\bar{M}^2(1+2\dz)$ are just the one-loop expansion of $2 \lambda V^2$ i.e the mass squared is defined as the coupling constant times the exact vev       squared,rather than the tree level vev squared, and the $\dz$  in(\ref{MMM}) is     absorbed in that redefinition of the $\ms$ mass. However, that definition of  the $\ms$ mass is generally gauge dependent when the gauge sector is included.   For the quartic \cc we have 
 \beqa
  \frac{\lambda^*}{\bar{\lambda}} \;\;&=& \;\;\frac{M^{*^2}\,\bar{v}^2}
{\bar{M}^2 \,v^{*^2}} 
  \;\;=\;\; \frac{M^{*^2}\,\zeta_v^{*^2}\,Z_{\phi}^*}{\bar{M}^2\,\bar{\zeta}_v^2
\,\bar{Z_{\phi}}} \;\;
=\;\; \frac{M^{*^2}}{\bar{M}^2}\;[1 \,+\, (\dZ_{\phi}^* \,-\, \bar{dZ}_{\phi})
 \,-\, 2 \bar{\dz}] \nonumber  \\
 &=&\;\;1 \;+\; \frac{\lambda}{16 \pi^2}[\,12 \lnM \,-\, 25 \,+\, 3\sqrt{3}\pi\, 
\;+\;{\cal O}(\frac{m^2}{M^2})]  \label{lamlam}
\eeqa
with $\dZ_{\phi}$ from (\ref{dz+}). Note the cancellation of $\dz$. 
The ratio $\lambda^*/\bar{\lambda}$ is independent of the choice of $\dz^*$ 
discussed below (\ref{delmu=}). An alternative calculation, which is 
manifestly independent of $\dz$,is 
\bb
 \frac{\lambda^*}{\bar{\lambda}} \;\;=\;\; \frac{\bar{Z_{\lambda}}}{Z_{\lambda}^*} 
 \;\;=\;\; \frac{\bar{Z_4}Z_{\phi}^{*^2}}{Z_4^* {\bar{Z_{\phi}}}^2} \;\;=\;\; 
 1 \;+\; (\bar{\dZ_4} \;-\; \dZ_4^*) \;+\; 2(\dZ_{\phi}^* \;-\;\bar{\dZ_{\phi}})                     \label{lamlam2} 
\ee
Again, keeping just the leading terms for $M^2 \gg m^2$, eqs (\ref{sigM}), 
(\ref{dz+}) and $\dz^* = 0$, give 
 \beqa
 \dZ_4^* \;&=&\; - \,3\, \lambda\, [\,\mbox{Re}I_{00}(M^2)\;+\; 3 I_{MM}(M^2)\,] 
\;+\;\frac{\lambda} {16 \pi^2}[-1] \nonumber  \\ 
 &=&\;\frac{\lambda}{16 \pi^2}\,\left[-\,3\left(-\Delta_{\epsilon}+ 
\lnM -2\right)-9\,\left(-\Delta_{\epsilon}+ \lnM -2 + \frac{\pi}
{\sqrt{3}}\right)-1\right]  \label{z4} 
\eeqa
Subtracting $\bar{\dZ_4}$ just removes the $\Delta_{\epsilon}$'s. Collecting  
the pieces in (\ref{lamlam2}) reproduces the result (\ref{lamlam}) \cite{sir}.

Proceeding to the fermion Green functions and \RCs, the terms in the Lagrangian 
(\ref{Lag}) quadratic in the fermion fields are 
\beqa
 {\cal L}_2 \;\;&=&\;\; \bar{\cal T}i \gamma \cdot\partial {\cal T}\;+\; 
\bar{\cal B}i \gamma \cdot \partial {\cal B}\;-
\;{{\cal Y}_0}_t \frac{{\cal V}}{\sqrt{2}} \bar{\cal T}{\cal T} \nonumber  \\
&=&\;\;\bar{t}(i\gamma\cdot \partial -m)t\;+\;\bar{b}i\gamma \cdot \partial b\;-\; 
(Z_3 \zeta_v -1)\,m\,\bar{t} t \nonumber \\
&+&\;\dZ_L \;\left(\,\bar{t}_L i \gamma \cdot \partial t_L \;+\;\bar{b}_L i 
\gamma \cdot \partial b_L\,\right) \nonumber \\
&+&\;\dZ_t \;\bar{t}_R i \gamma \cdot \partial t_R \;+\; \dZ_b \;\bar{b}_R i 
\gamma \cdot \partial b_R    \label{Lf2}
\eeqa
 Since we have only one nonzero Yukawa \cc and fermion mass, we usually       
suppress the subscript $t$: 
  $m_t \equiv m = {\cal Y}\frac{v}{\sqrt{2}}, \hspace{.5in} m_b = 0$. 
  This generates the fermion two-point \cts (for $-i\Sigma$):
 \beqa
   &t& : \;\;\;\;\;\;\;\; i\dZ_L\,\gamma p\,\frac{1-\gamma_5}{2} \;+\; i\dZ_t\,
\gamma p\,\frac{1+\gamma_5} {2} \;-\;i(Z_3 \zeta_v -1)\,m \label{ct2fa}  \\
 &b& : \;\;\;\;\;\;\;\;  i\dZ_L\,\gamma p\,\frac{1-\gamma_5}{2} \;+\; i\dZ_b\,
\gamma p\,\frac{1+\gamma_5}{2}  \label{ct2fb}
\eeqa
The vertex \cts generated by  substituting (\ref{renorm}), (\ref{vertrenorm}) 
into the trilinear terms in (\ref{Lag}) are displayed in fig. 3. 

\begin{center}\begin{picture}(300,100)(0,0)
%\SetScale{.8}

\CArc(35,60)(5,0,360)
\ArrowLine(15,30)(35,55)
\ArrowLine(35,65)(15,90)
\DashLine(40,60)(60,60){3}
\Text(36,60)[]{\Large \bf x}
\Text(5,10)[l]{$-i\delta Z_3 \frac{\cal Y}{\sqrt{2}} (1,i\gamma_5)$} 
\Text(11,95)[]{\Large t}
\Text(11,25)[]{\Large t}
\Text(65,60)[l]{\Large h,$\large \phi_0$}

\CArc(130,60)(5,0,360)
\ArrowLine(110,30)(130,55)
\ArrowLine(130,65)(110,90)
\DashLine(135,60)(155,60){3}
\Text(131,60)[]{\Large \bf x}
\Text(125,10)[l]{$i\delta Z_3{\cal Y}\frac{1+\gamma_5}{2}$} 
\Text(106,95)[]{\Large b}
\Text(106,25)[]{\Large t}

\CArc(215,60)(5,0,360)
\ArrowLine(195,30)(215,55)
\ArrowLine(215,65)(195,90)
\DashLine(220,60)(240,60){3}
\Text(216,60)[]{\Large \bf x}
\Text(210,10)[l]{$i\delta Z_3{\cal Y}\frac{1-\gamma_5}{2}$} 
\Text(191,95)[]{\Large t}
\Text(191,25)[]{\Large b}

\end{picture}\\
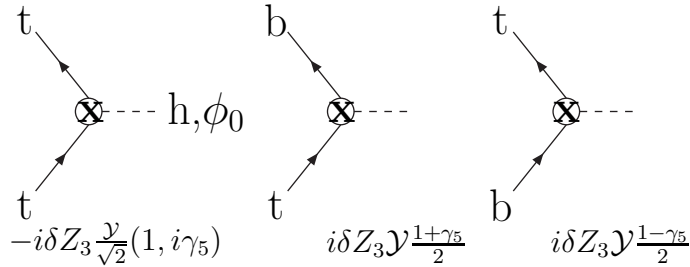
Fig.3. Yukawa vertex counter terms. 
\end{center}

The inverse of the complete renormalized propagator is
\bb
 S^{-1}(p) \;\;=\;\; \gamma p - m -\Sigma(p) \;\;=\;\; \gamma \,p\,
(1-A -B\gamma_5) \;-\; m\,(1 -C)    \label{sinv} 
\ee
Fig. 4 shows the one-loop Feynman diagrams and the \ct contributing to $\Sigma$: 

\begin{center}  \begin{picture}(440,50)(0,0)

\DashCArc(60,20)(20,0,180){6}
\Line(10,20)(110,20)
\Text(25,10)[]{\large t}
\Text(60,10)[]{\large t}
\Text(95,10)[]{\large t}

\DashCArc(180,20)(20,0,180){3}
\Line(130,20)(230,20)
\Text(145,10)[]{\large t}
\Text(180,10)[]{\large t}
\Text(215,10)[]{\large t}

\DashCArc(300,20)(20,0,180){3}
\Line(250,20)(350,20)
\Text(265,10)[]{\large t}
\Text(300,10)[]{\large b}
\Text(335,10)[]{\large t}

\CArc(395,20)(5,0,360)
\Line(370,20)(390,20)
\Line(400,20)(420,20)
\Text(397,20)[]{\Large \bf x}
\Text(375,10)[]{\large t}
\Text(415,10)[]{\large t}
\end{picture}\\
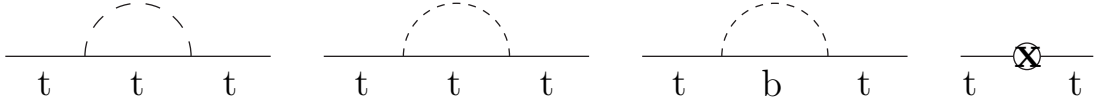
Fig.4.The one-loop self-energy diagrams of the t quark.
\end{center}
$\Sigma \;\;=\;\; \Sigma_{FD} \;+\; \Sigma_{ct}$: 
\beqa 
A \;\;&=&\;\; a-\frac{1}{2}\,\left(\dZ_L + \dZ_t\right), \nonumber  \\
B \;\;&=&\;\; b-\frac{1}{2}\,\left(-\dZ_L + \dZ_t\right), \nonumber  \\
C \;\;&=&\;\; c \;-\; \left(Z_3 \zeta_v -1 \right)  \label{ABC} 
\eeqa 
 with $a, b, c$ coming from the Feynman diagrams  $\Sigma_{FD}$. Rationalizing one 
obtains
\bb 
 S(p) \;\;=\;\; \frac{\gamma p(1-A-B\gamma_5) \;+\; m(1-C)}
{p^2((1-A)^2 -B^2)\;-\;m^2(1-C)^2}      \label{S} 
\ee
The mass shell condition for $p^2 = m^{*^2}$ is 
\beqa 
 0 \;\;&=&\;\; m^{*^2}((1-A)^2 -B^2) \;-\; m^2 (1-C)^2 \nonumber  \\ 
  \;\;&=&\;\; m^{*^2}(1 -2A) \;-\; m^2\,(1 - 2 C) \;+\; \mbox{two-loop} 
 \label{massshell} 
 \eeqa
The on-shell MOM \RC is then ($m^2 = m^{*^2}$) 
\bb
  A^* \;-\; C^* \;\;=\;\; 0   \label{AC}
\ee
The \ew contribution to $\frac{m^{*^2}}{\bar{m^2}}$ is 
\bb 
\frac{m^{*^2}}{\bar{m^2}} \;\;=\;\;1\;+\; 2\,(\bar{A} \;-\; \bar{C})\;\;=\;\; 
1 \;+\; 2\,(\bar{a} \,- \,\bar{c}) \;+\; 2 \bar{\dz}      \label{mm}  
\ee
where $\bar{a}$,$\bar{c}$ are just $a, c$ with the $\Delta_{\epsilon}$'s 
removed. The result of that calculation from the Feynman diagrams of fig. 4 
is \cite{BW} 
\beqa
 2(\bar{a}-\bar{c})\;\;&=&\;\;\frac{\yy}{16 \pi^2}\;\left[\,\Delta (r)\;+\;
\frac{3}{2}\lnm\, \right] \nonumber  \\
\Delta (r) \;\;&=&\;\; -\frac{1}{2} \;+\; 
\int_{0}^{1} dx\, (2-x)\;\ln(r^2(1-x)+x^2)  \label{Deltar} 
\eeqa
 with $r\;\equiv\;M/m$. There is also a one-loop QCD selfenergy diagram, 
to be added to the \ew diagrams of fig. 4, which adds  
\bb
 2(\bar{a} - \bar{c}) \;\;=\;\; 2\frac{\alpha_s}{\pi}(\frac{4}{3} \;-\; \lnm)  
   \label{QCD}     
\ee
Combining (\ref{mm}),(\ref{Deltar}),(\ref{QCD}) gives  \cite{BW}. See also  
\cite{HK}.
\bb
\frac{m^*}{\bar{m}}\;\;=\;\; 1\; +\;\frac{\alpha_s}{\pi}(\frac{4}{3}\;-\; \lnm) 
 \;+\; \frac{\yy}{32 \pi^2} \left[\;\Delta (r) \;+\; \frac{3}{2}\lnm\;\right]
\; +\; \bar{\dz}      \label{mmm}  
\ee
As in the bosonic case, the \MS\ mass depends on the \MS\ one-loop shift of 
the vev, $\dz$. (Again, as in the bosonic case, the $\dz$ in(\ref{mmm}) can be 
absorbed in the definition of $\bar{m}$, which is generally gauge dependent 
when the gauge sector is included).
 The $OS$ mass, $m^*$, which occurs in the $OS$ version of 
(\ref{massshell}) is the perturbative (all orders) pole mass - presumably        closely related to 
the reported experimental top quark mass. Also as in the bosonic case, we see 
from (\ref{ABC}) that $Z_3$ is fixed by the t mass \RC ($\zeta_v$ being already  
fixed), so $Z_3$ is not available to define the top Yukawa \cc as the value 
of some trilinear vertex function. Thus ${\cal Y}^*$ is fixed as 
\bb 
   {\cal Y}^* \;\;=\;\; \sqrt{2} \frac{m^*}{v^*}  \label{ystar} 
\ee

We relate the $MOM$ and \MS\ Yukawa \ccs.
\bb
\frac{{\cal Y}^*}{\bar{\cal Y}}\;\;=\;\;\frac{m^* \bar{v}}{\bar{m}v^*} \;\;=\;\;
 \frac{m^*}{\bar{m}} \left[\;1\;+\;\frac{1}{2}(\dZ_{\phi}^*-\bar{
\dZ_{\phi}})\;-\;\bar{\dz} \; \right]   \label{yy1} 
\ee
Again, as for the bosonic \cc, the $-\bar{\dz}$ in the brackets is cancelled 
by the $\bar{\dz}$ from $\frac{m^*}{\bar{m}}$. With $\dZ_{\phi}$ from 
(\ref{dz+}),this becomes 
\beqa
\frac{{\cal Y}^*}{\bar{{\cal Y}}} \;\;=\;\; 1\;&+&\; \frac{\alpha_s}{\pi}\, 
(\frac{4}{3}\;-\;\lnm) \;+\; \frac{\lambda}{32 \pi^2}\,[-1]\; \nonumber \\
&+&\;\; \frac{\yy}{32 \pi^2}\;\left[\,-\,\frac{1}{2}N_c\;+\;\Delta (r)\;+\;
(N_c\;+\;\frac{3}{2}) \,\lnm \; \right]   \label{yy2}   
\eeqa
Finally, 
\bb
 \lambda = \frac{\yy}{4}\frac{M^2}{m^2}
\ee
so the $\lambda$ term in (\ref{yy2}) may be replaced by $-r^2/4$ 
multiplied by $\yy/(32 \pi^2)$ giving \cite{BW} and see also \cite{HK}
\beqa
\frac{{\yy}^*}{{\bar{\yy}}^2}\;\;=\;\;1\; +\;2\frac{\alpha_s}{\pi}\,
\left(\,\frac{4}{3}-\lnm \,\right)  \nonumber \\
&+&\;\frac{\yy}{32 \pi^2}\,\left[\,-\frac{r^2}{2}\; +\; 
2\Delta (r) -N_c \;+\; (2N_c +3)\lnm\; \right]    \label{yy3}
\eeqa

We comment on the $\bar{\dz}$ appearing in the equation relating $m^*$ to 
$\bar{m}$ (\ref{mmm}). First note (\ref{zetabar}) that $\bar{\dz} 
\sim m^4/M^2 v^2 \sim {\cal Y}^4/\lambda$ can be sizable for a light Higgs 
boson.  
The top-quark two-point \ct (\ref{ct2fa}) may be rewritten  as
\bb
 - i \Sigma_{ct} \;\;=\;\; i\dZ_V\,(\gamma p \;-\; m)\;+\; i\dZ_A\gamma \,p 
\gamma_5  \;-\;i\dZ_m \,m      \label{zm}
\ee
where
\beqa
\dZ_V \;\;=\;\; \frac{1}{2}\,\left( \dZ_L \,+\, \dZ_t \right) \;,\nonumber  \\
\dZ_A \;\;=\;\; \frac{1}{2}\,\left(-\dZ_L\,+\,\dZ_t\right) \;,\nonumber  \\ 
\dZ_m \;\;=\;\; \dZ_y \;+\; \frac{1}{2}\dZ_{\phi} \;+\; \dz   \label{ZVA}  
\eeqa
This is convenient for on-shell $MOM$ renormalization, $\gamma p = m^*$ and 
$\dz^* = 0$. 
It is not suitable for \MS . $\dZ_m (\,\equiv \,\delta m/m)$ can not be 
treated as a \MS\ \rc\,, pure $\Delta_{\epsilon}$, because $\bar{\dz}$ is 
nonzero and not $\Delta_{\epsilon}$. In fact, $Z_m$ is not the renormalization 
of any parameter in the Lagrangian (\ref{Lag}). There is no fermion bare mass 
$m_0$.Thus changing $Z_m$ is not equivalent to a simple reparametrization of 
the Lagrangian. The Lagrangian does contain a bare Yukawa \cc, renormalized 
by $Z_y$: ${\cal Y}^*/\bar{{\cal Y}}\;=\;\bar{Z_y}/Z_y^*$.
So no $\dz$ appears in the equation relating ${\cal Y}^*$ to
$\bar{\cal Y}$ (\ref{yy2}).  

We conclude this section with an example showing the equivalence of a change 
of renormalization scheme and a reparametrization of the fields and parameters 
in the original Lagrangian. The example will be relevent to our dixcussion of 
the $\rho$ parameter in the next section. Consider 
\bb 
 \Delta \Sigma_{\phi}' \;\;=\;\; \Sigma_{\phi_0}'(0) \;-\; 
\Sigma_{\phi_+}'(0)      \label{delsig1}
\ee
with the $\Sigma_{\phi}'$ given in (\ref{sig+}),(\ref{sig0}). In the $MOM$      
renormalization scheme we can write 
\bb
 {\Delta \Sigma_{\phi}'}^* \;\;=\;\; N_c\, {\rm{y}}^{*^2}\, \left[\,1\; +\;  
\rm{y}^{*^2} \,(\rho_{FD}^{(2)}(r^*) \;+\; \rho_{ct}^{(2)^*}(r^*)) \;+\; 
\ldots\; \right]   \label{delsig2} 
\ee 
with $\rm{y}^2 \;\equiv\;\yy/(32 \pi^2)$.
The quantities $\rho^{(2)}$ are the results of a two-loop calculation \cite
{it},\cite{ftj}. (The one-loop result follows from (\ref{sig+}),(\ref{sig0}),
(\ref{dz+}),(\ref{dz0})). The \MS version is 
\bb
 \bar{\Delta\Sigma_{\phi}'} \;\;=\;\; N_c\,\bar{\rm{y}}^2\,\left[\,1\; +\; 
\bar{\rm{y}}^2(\rho_{FD}^{(2)}(\bar{r}) \;+\; \bar{\rho}_{ct}^{(2)}(\bar{r})) 
\;+\; \ldots \;\right]  \label{delsig3} 
\ee
To the two-loop order calculated, the $r^*$ and $\bar{r}$ in the $\rho^{(2)}$ 
functions and the $\rm{y}^{*^4}$ and $\bar{\rm{y}}^4 $ are equivalent. When 
we take the difference of the $MOM$ and \MS functions, 
the $\rho_{FD}^{(2)}$ terms drop out. 
\beqa
 \bar{\Delta\Sigma_{\phi}'}\; -\; {\Delta \Sigma_{\phi}'}^* \;\;&=&\;\; N_c \,
\rm{y}^4\, \left( \bar{\rho_{ct}^{(2)}}(r) \;-\;\rho_{ct}^{(2)^*}(r) \right)
 \nonumber  \\ 
 &=&\;\; N_c\,\rm{y}^4\, \left(2 \Delta (r) \;+\; 3 \lnm \right)  \label{delsig4}
\eeqa
 The explicit values of the $\rho_{ct}^{(2)}$ used to obtain the second line 
are discussed in section four.  $\Delta (r)$ is the 
function introduced in (\ref{Deltar}).
  
 We now check that this result,the \MS value of $\Delta\Sigma_{\phi}'$, is also
 obtained as a rescaling and reparametrization of ${\Delta \Sigma_{\phi}'}^*$. 
Since $\Sigma$ is essentially the inverse of the two-point function, the 
required rescaling factor is $ \bar{Z_{\phi}}/Z_{\phi}^{*}$ :  
\bb
 \frac{\bar{Z_{\phi}}}{Z_{\phi}^*}({\Sigma_{\phi_0}'}^*(0) \,-\,
 { \Sigma_{\phi_+}'}^*(0)) \;\;=\;\; (1\,-\,Z_1\rm{y}^2)N_c \bar{\rm{y}}^2 
\left[\,1\;+\;\rm{y}^2\,(Y_1 \;+\; \rho^{(2)^*}) \right]  \label{delsig5}
\ee
where
\beqa
  \frac{Z_{\phi}^*}{\bar{Z}_{\phi}} \;\;=\;\; 1\;+\;Z_1\rm{y}^2 
\;\;=\;\; 1\;+\;\rm{y}^2(-\frac{r^2}{2} \; -\;N_c \;+ \;2N_c\lnm) \label{Z1}  \\
  \frac{\rm{y}^{*^2}}{\bar{\rm{y}}^2} \;\;=\;\; 1\;+\;Y_1\rm{y}^2\;\;=\;\; 
1\; +\; \rm{y}^2(-\frac{r^2}{2} \;+\; 2 \Delta (r) -N_c \;+\; (2 N_c+3)\lnm) 
 \label{Y1}
\eeqa
Combining these expresions gives the result $(\ref{delsig5})=(\ref{delsig3})$ 

\section{Reduction formula}

A \RS refers to perturbative calculations of $\tau$ -functions.
\bb
 \tau \;\;=\;\; FT\langle T(\chi \ldots)\rangle   \label{tau}
\ee
The renormalization reparametrization (\ref{renorm}) of the original
Lagrangian (\ref{Lag}) generates in \pth `bare' feynman diagrams (no \cts,
but renormalized masses and \ccs), and \cts (Feynman diagrams including
$\dZ$ insertions).
\bb
  \tau \;\;=\;\; \tau_{FD} \;+\; \tau_{ct}   \label{tau2}
\ee
The $\tau_{FD}$ are the same function of the renormalized masses and \ccs in
any \RSs which introduces the same Z's (\ref{renorm}). There is implicit
scheme dependence in the definition of the renormalized masses and \ccs (\RCs).
The explicit scheme dependence is in the $\dZ$'s in the $\tau_{ct}$. To
one-loop order, the $\tau_{ct}$ are just $\dZ$ times a polynomial in momenta.
In higher orders, there are more complicated \cts arising from the nesting
of lower order \cts in lower order Feynman diagrams.

To connect a $\tau$-function to a physical observable (e.g. S-Matrix element)
requires use of the LSZ reduction formulas. As a preliminary, we note that
in all \RSs the physical mass $m^*$ is defined by the (perturbative) pole
of the complete renormalized two-point function. (In this formal perturbative
discusion we ignore difficulties with unstable and/or confined particles).
The distinction is between schemes which fix the residue of the pole to be
unity (*-schemes) and schemes which do not. The relation between the canonical
(bare) field $\chi_0$ and the various renormalized fields is
\bb
  \chi_0 \;\;=\;\; \sqrt{Z^*}\chi^* \;\;=\;\; \sqrt{\bar{Z}}\bar{\chi}  
\label{chi}
\ee
The LSZ reduction formula is simple in a *-scheme. (We show explicitly only
one external line factor).
\bb
 {\cal M} \;\;=\;\; {lim\choose q^2\rightarrow m^{*^2}}(q^2\;-\;m^{*^2})
 \tau^* \;\;=\;\; \underline{\tau^*}   \label{M}
\ee
where
\bb
 \tau = D \underline{\tau}   \label{amp}
\ee
D is the complete renormalized (e.g.$D^*$ or $\bar{D}$) two-point function, and
$\underline{\tau}$ is the fully amputated renormalized $\tau$-function.
\bb
 (q^2-m^{*^2})\bar{D} \;\;\ra \;\;\frac{Z^*}{\bar{Z}}   \label{Dbar}
\ee
But from (\ref{tau}),(\ref{chi}) and from (\ref{amp}),(\ref{Dbar})
\bb
 \tau^* \;\;=\;\; \sqrt{\frac{\bar{Z}}{Z^*}}\bar{\tau},\hspace{.5in} 
\underline{\tau^*}\;\;=\;\; \sqrt{\frac{Z^*}{\bar{Z}}} \underline{\bar{\tau}}  
\label{tautau}
\ee
Then
\bb
 {\cal M} ={\cal M}^* \;\;=\;\; \underline{\tau}^* \;\;=\;\; 
\sqrt{\frac{Z^*}{\bar{Z}} \underline{\bar{\tau}}} \;\;=\;\; \bar{\cal M}  
 \label{starbar}
\ee
$\tau^*$ is a function of the star renormalized masses and \ccs;
$\bar{\tau}$ is a function of the bar renormalized masses and \ccs.

\section{Applications}
\subsection{$H\;\ra\;f\,\bar{f}$ decay}

The first application is to the decay of a heavy Higgs to fermion anti-fermion.
Some of the contributiong Feynman diagrams and \cts are shown in fig5.  

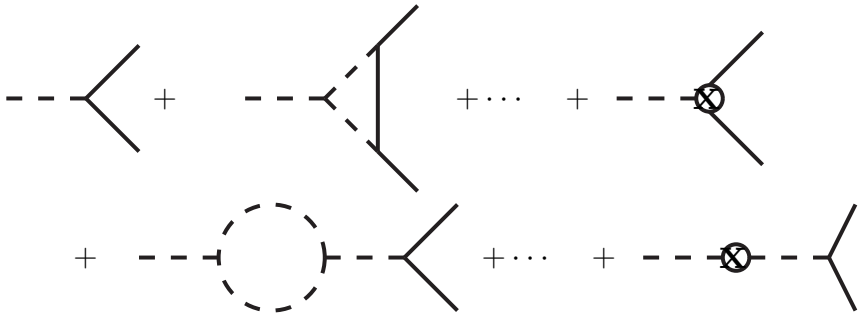
\begin{center} \begin{picture}(440,230)(0,0)
\SetWidth{1.5}

\DashLine(10,90)(40,90){6}
\Line(40,90)(60,110)
\Line(40,90)(60,70)

\Text(70,90)[]{$+$}

\DashLine(100,90)(130,90){6}
\DashLine(130,90)(150,110){6}
\DashLine(130,90)(150,70){6}
\Line(150,110)(165,125)
\Line(150,70)(165,55)
\Line(150,70)(150,110)

\Text(180,90)[l]{$+ \cdots \hspace{.2in} +$}

\DashLine(240,90)(270,90){6}
\CArc(275,90)(5,0,360)
\Line(275,95)(295,115)
\Line(275,85)(295,65)
\Text(275,90)[]{\Large \bf x}

\Text(40,30)[]{$+$}

\DashLine(60,30)(90,30){6}
\DashCArc(110,30)(20,0,360){6}
\DashLine(130,30)(160,30){6}
\Line(160,30)(180,50)
\Line(160,30)(180,10)

\Text(190,30)[l]{$+ \cdots \hspace{.2in} +$}

\DashLine(250,30)(280,30){6}
\CArc(285,30)(5,0,360)
\Text(285,30)[]{\Large \bf x}
\DashLine(290,30)(320,30){6}
\Line(320,30)(330,50)
\Line(320,30)(330,10)

\end{picture}\\
Fig.5. One-loop corrections to the amplitude for $H\rightarrow f\bar{f}$. In 
the heavy Higgs limit the boson loop diagram in the second line is dominant.
\end{center}

We consider the limit of a heavy Higgs, $M^2 \gg m^2$, $\lambda \gg\yy$. In our 
MOM scheme, with the Higgs field renormalized by the \GB field \rc, $Z_{\phi}$, 
and $\zeta_v=1$, the proper vertex Feynman diagrams and \ct in the first line of
 fig5 are all of order $\lambda \yy$, or $\lambda \frac{m}{M}$ times the tree 
term ($\sim {\cal Y}$). The only source of pure $\lambda$ one-loop corrections 
are the bosonic self-energy corrections to the external Higgs line. They 
survive LSZ amputation because $Z_{\phi}^* \neq Z_h^*$. (i.e. our $MOM $ 
scheme is 
not a *-scheme with respect to the Higgs field). Thus in the heavy 
Higgs limit, the leading terms in ${\cal M}$ are given by the external line 
factor times the tree term. The external line factor is determined by the 
analysis of the previous section (substitute $Z_h^*$ for $Z^*$ and $Z_{\phi}^*$ 
for $\bar{Z}$). 
\beqa
 {\cal M}\;\;&\doteq&\;\;\sqrt{\frac{Z_h^*}{Z_{\phi}^*}}{\cal M}_0^* \;\;=\;\; 
{\cal M}_0^* \left[1\;+\;\frac{1}{2}\,\left(\,\dZ_h^* \,-\, \dZ_{\phi}^*
\right)\;+\;\ldots\;\right]    \nonumber \\
&=&\;\; {\cal M}_0^* \,\left[\;1\;+\;\frac{1}{2}
\; \frac{\lambda^*}{16 \pi^2} (13 \,-\, 2\sqrt{3}\pi)\;+\;\ldots\; \right]  
\label{MEh}  
\eeqa
The last result is from (\ref{diffz}). This result was originally obtained 
by \cite{MW}. Recently the two-loop result has been calculated \cite{2lH}. 
\bb
 \Gamma \;\;=\;\;\Gamma_0^*\left\{\;1\;+\;2.12\frac{\lambda^*}{16\pi^2}
\;+\;\Gamma^{(2)^*}(\frac{\lambda^*}{16\pi^2})^2 \;+\;\ldots\;\right\}, 
\hspace{.3in} \Gamma^{(2)^*} \;\;=\;\; -32.66     \label{Gamstar}
\ee
We can use (\ref{lamlam}),$\lambda^*(\bar{\lambda})$, to rewrite 
(\ref{Gamstar}) in terms of the \MS \cc $\bar{\lambda}$. Setting the scale 
$\mu = M$, this gives 
\bb
 \Gamma \;\;=\;\; \Gamma_0^*\,\left\{1\;+\;2.12\,\frac{\bar{\lambda}}
{16\pi^2}\;-\;51.05\,(\frac {\bar{\lambda}}{16\pi^2})^2 \;+\;\ldots\;\right\}  
\label{Gam}
\ee
which makes the apparent convergence problem worse.

  Although (\ref{Gam}) is a legitimate reparametrization of (\ref{Gamstar})
(or (\ref{MEh})), it is not equivalent to a change of \RS from $MOM$ to \MS.
($\Gamma_0^*$ contains ${\cal Y}^{*^2}$)
The amputated \MS $\tau$-function, $\underline{\bar{\tau}}$, has no pure 
$\lambda$ term. Suppressing fermion spinors, 
\bb
  \underline{\bar{\tau}}(\bar{\cal Y},\bar{\lambda}) \;\;\doteq\;\; 
\frac{\bar{\cal Y}}{\sqrt{2}} \;+\; \bar{\lambda}(0 \;+\; {\cal O}(\yy) ;  
\label{tauhbar} 
\ee
The \MS calculation of ${\cal M}$, from (\ref{starbar}),(\ref{tauhbar}), is  
\bb
 \bar{{\cal M}} \;\;=\;\; \sqrt{\frac{Z^*_h}{\bar{Z}_h} \underline{\bar{\tau}}} 
\;\;\doteq\;\; \left\{\;1\;+\; \frac{1}{2}(\dZ_h^* \;-\; \bar{\dZ}_h)\;+\;\ldots
 \;\right\}\frac{\bar{{\cal Y}}}{\sqrt{2}}  \label{Mhbar1}
\ee
The $\bar{\dZ}_h$ just removes $\Delta_{\epsilon}$ from $\dZ_h^*$. 
So, with  $\dZ_h^*$ from (\ref{dzh}), this is
\bb
\bar{ {\cal M}} \;\;=\;\; \sqrt{\frac{Z^*_h}{\bar{Z}_h} \underline{\bar{\tau}}} 
\;\;\doteq\;\;  \frac{\bar{{\cal Y}}}{\sqrt{2}} \left[\;1\;+\;\frac{1}{2} \: 
\frac{\bar{\lambda}}{16 \pi^2}(12 \,-\, 2\sqrt{3}\pi)\;+\;\ldots\; \right]
 \label{Mhbar2}
\ee

We check that (\ref{Mhbar2}) is just the \MS reparametrization of 
${\cal M}^*$,   equation (\ref{MEh})
\bb
 {\cal M}^*\;\;\doteq\;\; \frac{{\cal Y}^*}{\sqrt{2}} \left\{1\;+\;\frac{1}{2}
 \: \frac{\lambda^*}{16 \pi^2}(13 \;-\; 2\sqrt{3}\pi) \;+\;\ldots \;\right\}  
\label{MEh2} 
\ee
In the heavy Higgs limit considered,
\bb
 {\cal Y}^* \;\;\doteq\;\; \bar{{\cal Y}} \;\left\{\;1\;+\;\frac{\lambda}
{16\pi^2}(-\frac{1}{2}\;+\;\ldots \;\right\}   \label{yy4}
\ee
Since the product ${\cal Y}\lambda$ is the same for star and bar to the 
order considered, this gives 
\bb
 {\cal M}^* \;\;\doteq\;\; \frac{\bar{{\cal Y}}}{\sqrt{2}}\;\left\{\;1\;+\;
\frac{1}{2} \: \frac{\bar{\lambda}}{16\pi^2}\; (12 \;-\; 2\sqrt{3}\pi)\;+\;
\ldots\;\right\} \;\;\doteq \;\; \bar{{\cal M}}     \label{Mhequiv} 
\ee
 
The  decay rate calculated from $\bar{{\cal M}}$ is
\bb
 \Gamma_{\bar{MS}} \;\;=\;\; \bar{\Gamma_0}\left\{\;1\;+\;1.12\,
\frac{\bar{\lambda}}{16\pi^2} \;+\;
 (?)(\frac{\bar{\lambda}}{16\pi^2})^2 \;+\; \ldots \;\right\}   \label{GamMS} 
\ee 
To complete the calculation of the coefficient of $\bar{\lambda}^2$ requires 
the two-loop $\lambda^2$ term in ${\cal Y}^*(\bar{{\cal Y}})$ which is not 
 available at the time of writing. But already at one-loop level the convergence looks better.

\subsection{$\rho$ - parameter}
Before moving on to the second application, we discuss anew the 
parameters in the reduced theory specified by the Lagrangian (\ref{Lag}).
The (bare) parameters in the Lagrangian are $\mu_0^2,\lambda_0,{\cal Y}_0$. 
After renormalization, (\ref{renorm}),these are replaced by $\mu^2,\lambda,
{\cal Y}$. The condition that the tree level vev of the shifted Higgs field 
vanish replaces $-\mu^2$ by $\lambda \; v^2$. In a MOM scheme, it is generally 
desirable to have the 'physical' masses as parameters, so we can trade 
$\lambda,{\cal Y}$ for $M,m$ via the relations $M^2=2\lambda \; v^2, 
m= {\cal Y}\;v/\sqrt{2}$, giving the set $\{v,M,m\}$. In $\overline{MS}$ 
schemes, it may be advantageous to use the coupling constants, which are 
directly related to bare parameters in the Lagrangian, rather than the 
masses, which are inextricably bound up with the vev. Then observables are
parametrized by the set $\{\bar{v},\bar{\lambda},\bar{{\cal Y}}\}$. 
In a MOM scheme,assuming that $M^*$ ($M_h$) and $m^*$ ($m_t$)  
 are measured (to some accuracy), we still have to know $v^*$ to give 
 a numerical value for observables such as  $\Gamma_{MOM}$ 
(\ref{Gamstar}). The determination of $v^*$ requires us to go beyond the 
reduced (``gaugeless'') theory we have been considering, and consider its 
embedding in the full standard Eletroweak theory.

In the full theory, $v^*$ is fixed by the relation
\bb
 v^{*^2} = 4 \frac{M_W^{*^2}}{g^{*^2}}   \label{vstar} 
\ee
 where $M_W^*$ is the W mass, and $g^{*^2}=e^{*^2}/s_w^{*^2}$ 
\bb
  e^{*^2} \;\;=\;\; 4\pi\alpha, \hspace{.5in}  
s_w^{*^2}\;\;=\;\;1\;-\;M_W^{*^2}/M_Z^{*^2}     \label{es} 
\ee
For precision \ew calculations, since $M_W$ is not so accurately known as 
$\alpha$ and $M_Z$, it is preferred to relate $v^{*^2}$ to the Fermi 
constant measured in $\mu$ decay. 
\bb 
 \frac{1}{2 v^{*^2}} \;\;=\;\; \frac{g^{*^2}}{8 M_W^{*^2}} \;\;=\;\; 
\frac{G_{\mu}}{\sqrt{2}} (1\;-\;\Delta r^*)     \label{vG} 
\ee
Here $\Delta r^*$ is the radiative correction to $\mu$ decay as originally 
computed by Marciano and Sirlin  (See Hollik \cite{Hollik} for a 
review in the context of the complete one-loop $MOM$ renormalization of 
the full \ew theory by BSH \cite{BSH}). 

At one-loop order, $\Delta r^*$ has quadratic dependence on $m_t$ ($\sim 
g^2 m_t^2/M_W^2 \sim m_t^2/v^2$), but only logarithmic dependence on $M_h$ 
(Veltman's screening theorem \cite{velt}), 
so in a calculation keeping only leading, quadratic in $M_h$, contributions, 
one can drop the $\Delta r^*$ term in (\ref{vG}) and replace ${\cal Y}^{*^2}$ 
and $\lambda^*$ in (\ref{Gamstar}) by $2\sqrt{2}G_{\mu}m_t^{*^2}$ and 
$G_{\mu}M_h^{*^2}/\sqrt{2}$ respectively . The interesting 
feature of (\ref{Gamstar}) is the quadratic and quartic  dependence on $M_h$ 
in one-loop and two-loop orders. Processes with an external Higgs line evade 
Veltman's screening theorem \cite{velt} that low energy processes with no 
external Higgs depend only logarithmically on the Higgs mass in one-loop 
order. For finite $M_h$, an accurate calculation requires additional       
contributions to (\ref{Gamstar}), some of which we have noted.(See fig. 5. See 
also \cite{Ri}). 
 
The second application involves the $\rho$ parameter. There are several 
definitions of $\rho$, all of which are unity at tree level in the standard 
\ew theory.\cite{vel} There are also different definitions of $\Delta\rho$, all of  
which provide a measure of the violation of the \ew i-spin symmetry 
produced by the unequal fermion Yukawa \ccs. The definition adopted by 
\cite{it},\cite{ftj} is that $\rho$ is the ratio of the effective low 
energy neutral current to charged current four-Fermi coupling constants. 
The effective low energy charged current four-Fermi \cc is the Fermi 
constant measured in $\mu$ decay (\ref{vG}). The effective low energy 
neutral four-fermi \cc is normalized by some function of the Weinberg 
angle such that the ratio is unity at tree level. Thus one could extract
$\rho$ from the ratio of the low energy cross sections 
for $\nu_{\mu}\;e$ to $\nu_{\mu}\;e$ ($Z$-exchange) and $\nu_{\mu}\;e$
 to $\mu\;\nu_e$ ($W$-exchange); see fig. 6.

\begin{center} \begin{picture}(440,200)(0,0)
\SetWidth{1.5}

\Vertex(30,60){5}
\Line(10,30)(50,90)
\Line(50,30)(10,90)

\Text(80,60)[]{$=$}

\ZigZag(120,60)(170,60){3}{6}
\Line(120,60)(100,90)
\Line(120,60)(100,30)
\Line(170,60)(190,90)
\Line(170,60)(190,30)

\Text(145,70)[]{\bf W}
\Text(145,50)[]{(\bf Z)}
\Text(100,20)[]{$\bf \nu_{\mu}$}
\Text(190,20)[]{\bf e}
\Text(100,100)[]{$\bf \mu(\nu_{\mu})$}
\Text(190,100)[]{$\bf \nu_e(e)$}

\Text(205,60)[]{$+$}

\ZigZag(240,60)(270,60){3}{4}
\Boxc(280,60)(20,20)
\ZigZag(290,60)(320,60){3}{4}
\Line(240,60)(220,90)
\Line(240,60)(220,30)
\Line(320,60)(340,90)
\Line(320,60)(340,30)

\Text(370,60)[]{$+ \hspace{.2in} \cdots$}
\end{picture}\\
\vspace{3cm}
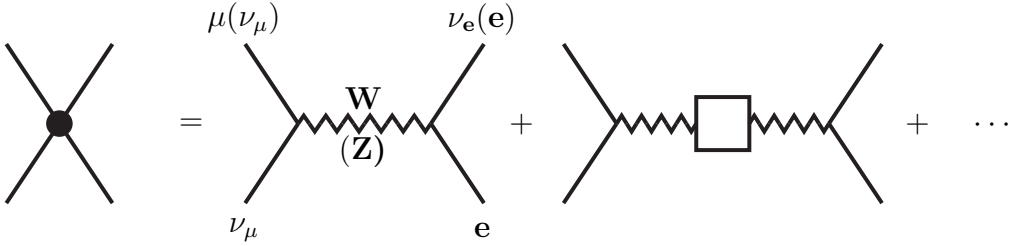
Fig.6.Radiative corrections to the $\rho$-ratio in the "gaugeless" limit 
  $m_H,m_t \gg m_W$.
\end{center}

 One can choose a \RS (either 
$MOM$ or \MS) in which the only source of contributions linear in $m_t^2$ or 
$M_h^2$ is selfenergy corrections to the exchanged vector boson propagator  
(fig. 6). Then
\bb
 \rho \;\;\doteq\;\; \frac{1}{c_w^2} \frac{D_Z(0)}{D_W(0)} \;\;=\;\; 
  \frac{1\;-\;\Pi_W(0)/M_W^2}{1\;-\;\Pi_Z(0)/M_Z^2}     \label{dotrho}
\ee
Here, $D, \Pi, M$ are all renormalized quantities, in whatever scheme, not 
yet specified. Again following \cite{it},\cite{ftj}, define 
\bb
 \Delta\rho \;\;=\;\; 1\;-\; \frac{1}{\rho}  \doteq \frac{P_Z\;-\;P_W}
{1 \;-\; P_W}        \label{Delrho} 
\ee
We have defined
\[
  P_{Z,W} \;\;\equiv\;\; \Pi_{Z,W}(0)/M_{Z,W}^2 
\]
The Feynman diagrams for the radiative corrections to $\mu$ decay,
$\Delta r$ in (\ref{vG}), are crossed versions of those for W exchange 
in fig. 6. Quadratic in $m_t$ contributions to $\Delta r$  come from 
W selfenergy insertions.
\bb
 \Delta r \;\;\doteq\;\; \frac{\Pi_w(0)}{M_W^2}   \label{dotr} 
\ee
Then, with the definition of $\Delta\rho$ (\ref{Delrho}),
\bb
 \Delta\rho \;\;\doteq\;\; \frac{1}{1 \;-\; \Delta r}(P_Z \;-\; P_W)  
\label{Delrho2} 
\ee

The gauge vector boson selfenergy functions $\Pi_{W,Z}$ are not included 
in the reduced ("gaugeless") theory we have considered. But in the full 
theory there are Slavnov-Taylor-Ward identites which relate the          
unrenormalized $\bf{\Pi}$,$\bf{\Sigma}$, for both W and Z.
\bb
 \frac{{\bf\Pi}_V(0)}{M_V^2} \;\;=\;\;\frac{{\bf\Sigma}_{\phi}(k^2)}{k^2}
\bigr |_{k^2 = 0} \; , \hspace{.3in}\mbox{ V=W,Z} \hspace{.2in} \phi\;\;=\;\; 
\phi_+,\phi_0    \label{stw} 
\ee
 The boldface $\bf{\Pi}$, $\bf{\Sigma}$ are given by the sum of        
 (regularized) Feynman diagrams with no \cts. In the detailed \RS of 
 \cite{BSH}, the $\gamma$,Z mixing lead to renormalized $\Pi$ functions 
which  have 
explicit dependence on $s_w$ (sine of the Weinberg angle), which the 
renormalized $\Sigma$ functions (\ref{sig+}),(\ref{sig0}) do not have.  
As already mentioned below (\ref{dz+}),(\ref{dz0}) when the full \ew 
theory is considered, one may prefer to fix $\delta Z$ by a different condition 
than (\ref{dz+}). A choice which leads to renormalized $\Sigma$ functions 
which satisfy S-T-W identites is given in App C. This does not have 
to be discussed here because the additional terms, dependent on the 
Weinberg angle, 
drop out in the difference between the charged and neutral functions (for those terms which could contribute 
quadratic in $m_t$ or $M_h$)
\beqa
  \frac{\Pi_Z(0)}{M_Z^2} \;-\; \frac{\Pi_W(0)}{M_W^2} \;\;\doteq\;\; 
\frac{{\bf\Pi}_Z(0)}    {M_Z^2} \;-\; \frac{{\bf\Pi}_W(0)}{M_W^2} \;\;=\;\; 
{\bf\Sigma}'_{\phi_0}(0) \;-\; {\bf\Sigma}'_{\phi_+}(0) \nonumber \\
= \;\; \Sigma'_{\phi_0}(0)+\dZ_{\phi} - \Sigma'_{\phi_+}(0)-\dZ_{\phi} 
\;\;=\;\; \Sigma'_{\phi_0}(0) \;-\; \Sigma'_{\phi_+}(0)    \label{pisig} 
\eeqa
These equations with (\ref{dz0}),(\ref{dz+}) give the one-loop result 
\bb
 \Delta\rho \;\;\doteq\;\;\Sigma'_{\phi_0}(0)\;-\;\Sigma'_{\phi_+}(0)\;\;=\;\; 
 \dZ_{\phi_0}\;-\;\dZ_{\phi_+} \;\;=\;\; N_c \;\frac{\yy}{32 \pi^2} 
  \label{rho1l}
\ee

Two groups \cite{it},\cite{ftj} have computed the two-loop \ew contributions 
to $\Delta\rho$ in $MOM$, using the STW relation (\ref{pisig}). 
(The authors of 
    the second reference have also directly calculated with the vector boson 
   selfenergies and obtained the same result, thus verifying (\ref{pisig}) 
   through two-loop order). The QCD correction to the one-loop \ew result 
 has been calculated in \cite{qcd}
\bb 
   \Sigma'_{\phi_0}(0) \;-\; \Sigma'_{\phi_+}(0)\;\;=\;\; N_c\,
\frac{{\cal Y}^{*^2}}{32\pi^2} \,   \left\{\,1 \;-\;\frac{2}{9}\frac{\alpha_s}
{\pi}(\pi^2 + 3)  
\;+\;\frac{{\cal Y}^{*^2}}{32\pi^2}\rho^{(2)^*}(r) \,\right\}  \label{rho2star} 
\ee
\bb
\Delta \rho =\;\;\frac{1}{1-\Delta r^*} N_c\,
\frac{{\cal Y}^{*^2}}{32\pi^2} \,   \left\{\,1 \;-\;\frac{2}{9}\frac{\alpha_s}
{\pi}(\pi^2 + 3)
\;+\;\frac{{\cal Y}^{*^2}}{32\pi^2}\rho^{(2)^*}(r) \,\right\}  \label{star2}
\ee
By (\ref{ystar}), (\ref{vG}),
\bb
 \frac{{\cal Y}^{*^2}}{1 -\Delta r^*}\;\; =\;\; 2\sqrt{2}\,G_{\mu}\,m^{*^2} 
      \label{y,G} 
\ee
which is the motivation for the definition of $\Delta\rho$ in (\ref{Delrho}). 
From (\ref{Delrho2}),(\ref{pisig}),(\ref{rho2star}),(\ref{y,G}), 
 we arrive at  
\beqa 
\Delta\rho\;\;&=&\;\;N_c\,x_t^*\;\left\{\;1\;-\;\frac{2}{9}\,\frac{\alpha_s}{\pi}
\,(\pi^2\,+\,3)\;+\;x_t^*\,\rho^{(2)^*}(r)\;+\;\ldots \;\right\}, \nonumber \\
 x_t^*\;\;&\equiv&\;\;\frac{G_{\mu}m_t^{*^2}}{8\sqrt{2}\pi^2}. \label{rho2star2} 
\eeqa
All of the \ew quantities in this formula are computed in MOM, so this 
$\Delta \rho$ is $\Delta \rho_{MOM}$.
The small $r$ and large $r$ behaviors of $\rho^{(2)^*}(r)$ are \cite{it},
\cite{ftj} 
\beqa
\rho^{(2)^*}(r)\;\; &\simeq&\;\;-2\pi^2\;+\;19 \hspace{1.75in} (r\rightarrow 0) \\
\rho^{(2)^*}(r)\;\; &\simeq& \;\;\frac{3}{2}(\ln r^2)^2 \;-\;\frac{27}{2}
\ln r^2\;+\;\pi^2\;+\;\frac{49}{4} \hspace{.2in} (r \gg 1)   \label{rho2star3}
\eeqa
So for the $MOM$ calculation of $\Delta\rho$, the large $m_t$ behavior is 
of order $G_{\mu}\,m_t^2$ and $(G_{\mu}\,m_t^2)^2$, and the large $M_h$ 
behavior is order $O\left((G_{\mu}m_t^2)^2 (\ln M_h^2/m_t^2)^2\right)$.

We now consider some reparametrizations of this result. 
First, we collect the transformation equations for the relevent parameters.
Copying (\ref{mmm}),(\ref{yy2})
\bb
\frac{m^*}{\bar{m}}\;\;=\;\; 1\; +\;\frac{\alpha_s}{\pi}(\frac{4}{3}\;-\; \lnm)
 \;+\; \frac{\yy}{32 \pi^2} \left[\;\Delta (r) \;+\; \frac{3}{2}\lnm\;\right]
\; +\; \bar{\dz}      \label{mmmm}
\ee
\beqa
\frac{{\cal Y}^*}{\bar{{\cal Y}}} \;\;=\;\; 1\;&+&\; \frac{\alpha_s}{\pi}\,
(\frac{4}{3}\;-\;\lnm) \;+\; \frac{\lambda}{32 \pi^2}\,[-1]\; \nonumber \\
&+&\;\; \frac{\yy}{32 \pi^2}\;\left[\,-\,\frac{1}{2}N_c\;+\;\Delta (r)\;+\;
(N_c\;+\;\frac{3}{2}) \,\lnm \; \right]   \label{yy22}
\eeqa 
and, see (\ref{lamlam})
\beqa 
 \frac{v^*}{\bar{v}} &=& 1+\frac{1}{2}(\delta \bar{Z}_{\phi}-\delta Z^*_{\phi})
+ \delta \bar{\zeta}_v  \nonumber  \\
  &=& 1 + \frac{\lambda}{32 \pi^2}[1] +N_c\frac{\yy}{32 \pi^2}[-\ln{\frac{m^2}
{\mu^2}}+\frac{1}{2}] + \delta \bar{\zeta}_v   \label{vv} 
\eeqa

The first reparametrization  is just to transform from "on-shell" top mass       $m_t^*$ to the \MS mass in (\ref{rho2star2}). 
This produces  
\beqa
\Delta\rho \;\;=\;\; N_c\;\bar{x_t}\;\{\;1\;&-&\;\frac{2}{9}\;
\frac{\alpha_s}{\pi}\,\left[\pi^2\,-\,9\,+\,9\lnm\,\right] \nonumber  \\
&+&\;\bar{x_t}\,\left[\,\rho^{(2)^*}(r)\;+\;2 \Delta (r) +3\lnm\,\right]
\;+\;2\,\bar{\dz} \;\}   \label{rho2bar1a}   \\
\bar{x_t}\;\;&\equiv&\;\;\frac{G_{\mu}{\bar{m_t}}^2}{8\sqrt{2}\pi^2} \label{rho2bar1b} 
\eeqa
To two-loop order, all the quantities $(m,\;r=M/m)$ may be taken to be 
$\overline{MS}$ quantities. Thus this $\Delta \rho$ is a candidate for 
$\Delta \rho_{\overline{MS}}$.
As discussed at the end of section two, reparametrization of the quark mass 
in the \ew theory is questionable because there is no bare mass parameter in 
the Lagrangian; and it leads to the appearance of the singular $\bar{\dz}$ in 
(\ref{rho2bar1a}). This suggests that a better reparametrization would be of 
the Yukawa \cc.\cite{BW} There is a bare Yukawa \cc in the Lagrangian and the   
transformation from ${\cal Y}^*$ to $\bar{\cal Y}$ (\ref{yy22}) does not 
involve $\bar{\dz}$. Substitution of (\ref{yy22}) into (\ref{star2}) yields 
\beqa
\Delta\rho\;\;=\;\;\frac{1}{1-\Delta r^*}N_c\;\bar{y}^2\;\{\;1\;&-&\;
\frac{2}{9}\,\frac{\alpha_s}{\pi}\,\left[\,\pi^2\,-\,9\,+\,9\lnm \,\right] 
\nonumber \\ 
&+&\;\bar{y}^2 \;\left[\,\rho^{(2)^*}-\,\frac{r^2}{2}\;+\;
2\Delta(r)\;-\;N_c\;+\;(2 N_c +3)\, \lnm \,\right] \,\}  \label{rhobb} 
\eeqa
$$y^2\;=\;\frac{{\cal Y}^2}{32 \pi^2} $$

There are two very interesting features of this parametrization. If we 
set the scale mass $\mu$ equal to $m_t$, the coefficient of $\alpha_s$ 
becomes very small \cite{Jer}and the two-loop \ew correction to the one-loop     result
 becomes larger than the QCD correction. The second is the appearance of 
the $-r^2/2$, which comes from the scalar \cc contribution to ${\cal Y}^*$ 
($\bar{\cal Y}$), (\ref{yy2}). Thus the large $M_h$ behavior in this 
parametrization is order $G m_t^2 G M_h^2$.

However, the presence of the factor $1/(1-\Delta r^*)$ implies that this is a 
parametrization in a mixed scheme.
In appendix A we show that
\bb
\frac{1}{1 -\overline{\Delta r}} \; \frac{1}{\bar{v}^2}\;\; \doteq  \;\;
\frac{1}{1 -\Delta r^*} \; \frac{1}{v^{*^2}} \;=\;\sqrt{2}G_{\mu}  \label{rv}
\ee 
From which follows
\bb
\frac{1}{1-\Delta r^*}\;=\;\frac{v^{*^2}}{\bar{v}^2}\;\frac{1}{1-\Delta \bar{r}}
 =\frac{1}{1-\Delta \bar{r}}\{1+N_c y^2[\frac{1}{2}r^2+
N_c (1-2\lnm )]+ 2 \delta \bar{\zeta}_v  \; \}  \label{denom}
\ee
Substitution of (\ref{denom}) into  (\ref{rhobb}) yields
\beqa
\Delta\rho\;\;=\;\;\frac{1}{1-\Delta \bar{r}}N_c\;\bar{y}^2\;\{\;1\;&-&\;
\frac{2}{9}\,\frac{\alpha_s}{\pi}\,\left[\,\pi^2\,-\,9\,+\,9\lnm \,\right]
\nonumber \\
&+&\;\bar{y}^2 \;\left[\,\rho^{(2)^*}+\;
2\Delta(r)\; +3\,\lnm \,\right] \,+ 2\;\delta\bar{\zeta}_v\;\}  \label{rhoms} 
\eeqa
 But from (\ref{delsig3})
\bb
 \rho^{(2)^*}\;+\;2\;\Delta(r)+3\;\lnm\;=\;\bar{\rho}^2  \label{eq}
\ee
and by (\ref{rv})
\bb
 \frac{\bar{y}^2}{1-\Delta\bar{r}}\;=\;\bar{x}_t   \label{eq2}
\ee
With these relations,this $\Delta\rho$ from (\ref{rhoms}) is identical to 
$\Delta\rho_{\overline{MS}}$ from (\ref{rho2bar1a}).

These alternative parametrizations of $\Delta\rho$ may be used as one
estimate of the error arising from truncation of the perturbation series
at finite (in this case, two-loop) order. Since these are just
reparametrizations, exact (all orders) calculations, using exact (all
orders) relations between the parameters, must give the same numerical
result. When the exact result for two different parametrizations is separated
into a finite order calculated
part plus uncalculated remainder, it follows that at least one of the
remainders is the same order of magnitude as the difference of the two
finite order calculated terms.

Because of the explicit $\ln{\mu^2}$ dependence in the formulas (\ref{mmmm})
to (\ref{eq}), and the implicit dependence on $\ln{\mu^2}$ in choice of
value for $\alpha_s$ which appears in these formulas, the questions of
truncation error and scheme dependence become entangled with questions of
scale dependence. Thus there are many choices to make as to quantities to
compare. We have chosen to compare $\Delta\rho_{MOM}\;$(\ref{rho2star2}),
and the alternative parametrizations we have called $\Delta\rho_{mix}\;$
(\ref{rhobb}), and $\Delta\rho_{\overline{MS}} \;$(\ref{rho2bar1a}), or
(\ref{rhoms}). And we make these comparisons for three choices of scale(s).
First, we take all explicit $\mu s$ equal to $M_W$, but $\alpha_s$ at
scale of $m_t$. Second, we take all $\mu$ s which come from QCD to be
$m_t$, and all $\mu$ s from weak interactions to be $M_W$. Third, we take 
all $\mu$ s equal to $m_t$. The results 
are given in Tables 1,2,3. \vspace{.3in} 

Table 1. Two-loop values of $\Delta\rho$ in different parametrizations. \\
\hspace*{.5in}$\mu = M_W.\;\; m_t = 180. \;\; r = M_h/180. \;\; \alpha_s (m_t) =  .107.$  \vspace{.1in} 

\begin{tabular}{|c|c|c|c|c|c|} \hline
\multicolumn{1}{|c|}{ }  &
\multicolumn{5}{c|}{$10^3 \; \Delta\rho$} \\ \cline{2-6}
$M_h$ & MOM & MIX & diff & $\overline{MS}$ & diff \\  \hline
60  & 9.05 & 9.07 & -.02 & 7.76 & 1.29 \\
150 & 8.96 & 8.99 & -.03 & 9.00 & -.04 \\
300 & 8.87 & 8.91 & -.04 & 8.82 & .05  \\
600 & 8.80 & 8.84 & -.04 & 7.36 & 1.44  \\
1000 & 8.77 & 8.78 & -.01 & -8.09 & 16.86  \\  \hline
\end{tabular}   \vspace{.3in}

Table 2. Two-loop values of $\Delta\rho$ in different parametrizations. \\
\hspace*{.5in}$\mu_{QCD} = m_t.\;\;\mu_w = M_W. \;\; m_t = 180.$ \vspace{.1in}

\begin{tabular}{|c|c|c|c|c|c|} \hline
\multicolumn{1}{|c|}{ }  &
\multicolumn{5}{c|}{$10^3 \; \Delta\rho$} \\ \cline{2-6}
$M_h$ & MOM & MIX & diff & $\overline{MS}$ & diff \\  \hline
60  & 9.05 & 9.06 & -.01 & 7.27 & 1.78 \\
150 & 8.96 & 8.97 & -.03 & 8.91 & .05 \\
300 & 8.87 & 8.89 & -.02 & 8.92 & -.05  \\
600 & 8.80 & 8.84 & -.04 & 8.28 & .52  \\
1000 & 8.77 & 8.86 & -.09 & -2.26 & 11.03  \\  \hline
\end{tabular}   \vspace{.3in}  

\clearpage

Table 3. Two-loop values of $\Delta\rho$ in different parametrizations. \\
\hspace*{.5in}$\mu = m_t. \;\; m_t = 180.$ \vspace{.1in}

\begin{tabular}{|c|c|c|c|c|c|} \hline
\multicolumn{1}{|c|}{ }  &
\multicolumn{5}{c|}{$10^3 \; \Delta\rho$} \\ \cline{2-6}
$M_h$ & MOM & MIX & diff & $\overline{MS}$ & diff \\  \hline
60  & 9.05 & 9.12 & -.07 & 4.60 & 4.45 \\
150 & 8.96 & 9.04 & -.08 & 8.94 & .02 \\
300 & 8.87 & 8.97 & -.10 & 8.93 & -.06  \\
600 & 8.80 & 8.90 & -.10 & 8.71 & .09  \\
1000 & 8.77 & 8.88 & -.11 & 4.28 & 4.49  \\  \hline
\end{tabular}  \vspace{.3in}  

  From the tables we see that for values of $r = M_h/m_t$ departing 
significantly from one, at least one of the truncated perturbation 
series for $\Delta\rho$ is very bad.   The 'visible' problem is coming from the $\delta\
\bar{\zeta}_v$ which appears in the parametrization we have called 
$\Delta\rho_{\overline{MS}} \;$ (\ref{rho2bar1a}), (\ref{rhoms}),      and which blows 
up as $1/r^2$ or as $r^2$ as $r$ goes to zero or infinity. This supports 
the contention that it is better to transform the Yukawa coupling 
constant than the quark mass. Of course we should not be surprised that 
the perturbation theory has failed for a Higgs mass of order one TeV. It 
has been long known that partial wave unitarity is violated by the 
tree level perturbative amplitudes for a Higgs of this mass. In fact, 
despite the small differences between the truncated perturbative results 
for $\Delta\rho_{MOM}$ and $\Delta\rho_{MIX}$, we should expect both 
of these perturbative expansions to be bad for a Higgs mass of one TeV.
(The smallness of the difference of the truncated expansions is a 
necessary, but not sufficient, condition for both of the expansions to 
be good). All we can say is that for Higgs mass in the range of one 
hundred to six hundred GeV, these results are consistent with the    
two-loop results for  $\Delta\rho_{MOM}$ and $\Delta\rho_{MIX}$ being 
accurate to the order of one percent. 

   For an alternative and more detailed discussion, with perhaps some 
difference of interpretation, we refer to Kniehl and Sirlin \cite{KS}.

\section{Appendix A}

In this appendix we neglect numerical factors $32\pi^2$,$\sqrt{2}$ 
(e.g.absorb them in definitions of ${\cal Y}$ and G.) Then (\ref{y,G}) is 
written as
\bb
 \frac{1}{1 \;-\;\Delta r^*}{\cal Y}^{*^2}\;\;=\;\; G_{\mu}m^{*^2}  \label{a1}
\ee
and (\ref{dotr}) is
\bb
   \Delta r \;\;\doteq\;\; \frac{\Pi_W(0)}{M_W^2}      \label{a2}
\ee
Then
\bb
M_W^{*^2}\;\left(\,1\;-\;\Delta r^*\,\right)\;\;\doteq \;\; M_W^{*^2}\;- 
\;\Pi_W^{*^2}(0)      \label{a3}
\ee
  (Suppress the subscript W)
\bb
  \Pi(k^2) \;\;=\;\; {\bf\Pi}(k^2) \;+ \;\delta M^2 \;+ \;(k^2\,-\,M^2)\,\delta Z   
\label{a4}
\ee
in either MOM or \MS. Then
\beqa
\delta M^{2^*} \;\;&=&\;\; -{\bf\Pi}(M^{*^2})   \\
0 \;\;&=&\;\; M^{*^2} \;-\; \bar{M}^2 \;+\; \bar{\Pi}(M^{*^2}) \nonumber  \\
\bar{\Pi}(M^{*^2}) \;-\; \bar{\Pi}(0) \;\;&=&\;\; {\bf\Pi}(M^{*^2}) \;+\;
 M^{*^2}\,\delta\bar{Z}\;-\;{\bf\Pi}(0)  \nonumber \\   
 &=& \;- \Pi^*(0) + (\delta\bar{Z}\;-\;\delta Z^*)M^{*^2}   \label{a5}
\eeqa
Then
\beqa
\bar{M}^2\;\left(\,1\;-\;\overline{\Delta r}\right) \;\;&=&\;\; \bar{M}^2 
\;-\; \bar{\Pi}(0) \;\;= \;\;M^{*^2}\;+\;\bar{\Pi}(M^{*^2})\;-\;\bar{\Pi}(0) 
\nonumber \\
 &=&\;\; M^{*^2} \;-\; \Pi^*(0) \;+\;\left(\delta\bar{Z} \;-\;\delta 
Z^*\right)\;M^{*^2}    \label{a6}
\eeqa
Then
\beqa
  \setlength{\jot}{.4in}
 \frac{\bar{g}^2}{\bar{M}_W^2} \; \frac{1}{1-\overline{\Delta r}}
  \;\;=\;\; \frac{\bar{g}^2}{\bar{M}_W^2 \;-\; \bar{\Pi}_W(0)} \;\;=\;\;
 \frac{g^{*^2}(1\;+\;2\,\left(\delta Z_g^* -\delta\bar{Z}_g)\right)}
  {M_W^{*^2}\,\left(1\;-\;\Pi_W^*(0)/M_W^{*^2} \;-\; \left(\delta Z_W^* 
\;-\;\delta\bar{Z}_W\right)\right)} \nonumber  \\
  =\;\; \frac{g^{*^2}}{M_W^{*^2}} \; \frac{1}{1\,-\,\Delta r^*}\; \left[\, 
1\;+\;2\, \left(\delta Z_g^* \; -\;\delta\bar{Z}_g \right) \;+\;
\delta Z_W^* \;-\;\delta\bar{Z}_W \;+\; \mbox{two-loop} \right]          \label{a7}
\eeqa
The combination of \rcs which appears here is precisely the combination which
enters into the renormalization of the proper vertex function.
($2(\delta Z_1^W \;-\; \delta Z_2^W)$) in the notation of eq. (4.10) of          \cite{Hollik}). In

the $MOM$ scheme of BSH \cite{BSH}, this combination has no contribution linear
in $m_t^2$, so we can write
\bb
 \frac{1}{1\;-\;\overline{\Delta r}} \; \frac{1}{\bar{v}^2} \;\;\doteq\;\;
   \frac{1}{1\;-\;\Delta r^*} \; \frac{1}{v^{*^2}}    \label{a8}
\ee
\section{Appendix B}

At the end of section one (see \ref{delsig4}) we made use of the fact that
the difference of the two-loop MOM and \MS contributions to $\Delta\rho$ is
just equal to the difference of the corresponding two-loop \cts.
\bb
  \bar{\rho}^{(2)} \;-\; \rho^{*(2)} \;\;=\;\; \bar{\rho}_{ct}^{(2)}\; -\; 
\rho_{ct}^{*(2)}     \label{b1}
\ee
We have independently computed these \cts.
\bb
 \rho_{ct}^{*(2)}\;\;=\;\; 3\Delta_{\epsilon}-3\lnm \;-\; 2\Delta (r)\;-\;
\frac{3}{2},   \hspace{.4in}  \bar{\rho}_{ct}^{(2)} \;\;=\;\; 
3\Delta_{\epsilon} \;-\;\frac{3}{2}    \label{b2}
\ee
One can readily verify that $\rho_{ct}^{*(2)}$ given here (see (\ref{Deltar})
is identical to eq (11) of the second FTJ paper \cite{ftj}. However FTJ have
omitted the $-3/2$ from the \MS \ct. But the origin of this term is clear. It
is the product of $1/\epsilon$ from one-loop $\delta Z$ times a term of order
$\epsilon$ from the dimensionally regulated one-loop integral in which the
$\delta Z$ \ct is embedded. Since the divergent part of $\delta Z$ and the
"bare" one-loop dimensionally regulated Feynman integrals are the same for
MOM and \MS, the same $-3/2$ occurs in both \cts. It is only at one-loop
order, where the \ct is pure $\delta Z$, that the \MS \ct only subtracts
$1/\epsilon$ from the Feynman integrals. Finally, note that the example
given at the end of section two verifies the difference of these two \cts.
(This point has also been commented on by Kniehl and Sirlin \cite{KS}).

\section{Appendix C}

For the full Electroweak theory we follow generally the renormalization 
prescriptions of \cite{BSH},\cite{Hollik}. In the vector boson sector, they 
impose five \RCs: the renormalized $M_W,M_Z$ are the physical masses. The 
photon mass is zero, and the residue of the pole of the renormalized 
photon two-point function is unity. The mixed $\gamma - Z$ two-point 
function is zero at $k^2 = 0$.  There are only four renormalization 
constants: $Z_W,Z_B,Z_g,Z_{g'}$. So $Z_{\phi}$ is also used to enforce 
these conditions. The resulting $\delta Z_{\phi}$, proportional to the 
top Yukawa coupling constant squared, is
\bb
 \delta Z_{\phi} \doteq x \; [-\Delta_{\epsilon} + \ln{\frac{m^2}{\mu^2}} + 
\frac{1}{2} (\frac{c^2}{s^2} -1)]  \label{c1} 
\ee
\bb
  x = N_c \; \frac{{\cal Y}^2}{16 \pi^2}   \label{c2}  
\ee 
Substituting this $\delta Z_{\phi}$ into (\ref{sig+}),(\ref{sig0}), we 
find for the terms proportional to x,
\bb
  \Sigma'_{\phi_+} \doteq \; -x[\frac{1}{2}\;\frac{c^2}{s^2}]  \hspace{1.5in} 
 \Sigma'_{\phi_0} \doteq \; -x[\frac{1}{2}\;\frac{c^2-s^2}{s^2}]  
 \label{c3} 
\ee
The terms proportional to the top mass squared in the vector boson selfenergy 
functions are 
\bb 
 \frac{\Pi_W (0)}{M_W^2} \doteq \: N_c \frac{g^2}{16 \pi^2}\frac{m^2}{M_W^2}[
\frac{-1}{4}\:\frac{c^2}{s^2}], \hspace{1 in}  \frac{\Pi_Z (0)}{M_Z^2} \doteq 
\; N_c \frac{g^2}{16 \pi^2}\frac{m^2}{M_Z^2}[\frac{-1}{4}\;\frac{c^2-s^2}{
s^2}]    \label{c4}  
\ee
Using 
$$ {\cal Y}^2= g^2\frac{m^2}{2\;M_W^2}, \hspace{1in} M_Z^2\;c^2 = M_W^2 $$
we see that the S-T-W relations are satisfied by both the charged and neutral 
functions and the differences are independent of the Weinberg angle.

\end{document}